\documentclass[prb,twocolumn,amsmath,amsfonts,10pt]{revtex4-1}
\usepackage{epsf}
\usepackage{graphicx}           
\usepackage{color}

\usepackage{amssymb}
\usepackage{amsmath}
\usepackage{revsymb4-1}
\usepackage{rotating}
\usepackage{tabulary}
\usepackage{setspace}
\usepackage{multirow}
\usepackage{color}
\usepackage{wasysym}
\definecolor{green0}{rgb}{0,0.5,0}
\usepackage{bm}
\usepackage{fixmath}
\usepackage{amsbsy}
\usepackage{pifont}
 \definecolor{darkgreen}{rgb}{0,0.5,0} % VM
 \usepackage{color}
 \usepackage[normalem]{ulem}
 \usepackage{cancel}
\newcolumntype{C}{>{\centering\let\newline\\\arraybackslash\hspace{0pt}}m{0.8cm}}
\newcolumntype{D}[1]{>{\centering\let\newline\\\arraybackslash\hspace{0pt}}m{#1}}
\newcolumntype{E}{>{\centering\let\newline\\\arraybackslash\hspace{0pt}}m{0.9cm}}
 \newcommand{\down}{{\bf --} \hskip -0.31cm $\downarrow$ }
 \newcommand{\up}{{\bf --} \hskip -0.31cm $\uparrow$ }
 \newcommand{\zero}{{\bf --} \hskip -0.31cm \phantom{$\uparrow$} }
 \newcommand{\double}{{\bf --} \hskip -0.364cm  $\uparrow \negthickspace \downarrow$ }

\newcommand{\ups}{{$|\uparrow\rangle_i$}}
\newcommand{\downs}{{$|\downarrow\rangle_i$}}

 \definecolor{darkgreen}{rgb}{0,0.5,0} % VM
 \usepackage{color}
 \usepackage[normalem]{ulem}
 \usepackage{cancel}
\begin{document}          

\title{Entanglement patterns and generalized correlation functions\\ in quantum many-body systems}                                                        

%RevTeX form:
\author{G. Barcza$^{1}$, R. M. Noack$^2$, J. S\'olyom$^1$, \"O.~Legeza$^{1}$} 
\affiliation{
   $^1$ Strongly Correlated Systems ``Lend\"ulet'' Research group,
   Wigner Research Centre for Physics, H-1525 Budapest, Hungary\\ 
   $^2$ Fachbereich Physik, Philipps-Universit\"at Marburg, 35032 Marburg, Germany}

\date\today

%\vskip -8pt
\begin{abstract}
We introduce transition operators that in a given basis of the
single-site states of a many-body system have a single non-vanishing 
matrix element and introduce their correlation functions. 
We show that they fall into groups that decay with the same rate. 
The mutual information defined in terms of the von Neumann entropy between
two sites is given in terms of these so-called generalized correlation 
functions.
We confirm numerically that the long-distance decay of the mutual
information follows the square of that of the most slowly decaying
generalized correlation function.
The main advantage of our procedure is that,
in order to identify the most relevant physical processes,
there is no need  
to know {\it a priori} the nature of the ordering in the 
system, i.e., no need to
explicitly construct particular physical correlation functions.
We explore the behavior of the mutual information and the generalized
correlation functions
for comformally invariant models  
and for the SU($n$) Hubbard model with $n=2,3,4$, and 5, which are, 
in general, not conformally invariant.
In this latter case, we show that 
for filling $f=1/q$ and $q<n$, the ground state consists of
highly entangled $q$-site units that are 
further entangled by single bonds. 
In addition, we extend the picture of the two-site mutual information
and the corresponding generalized correlation functions to the
$n$-site case.

%\PACS{(75.10.Jm)}
\end{abstract}

\pacs{PACS number: 75.10.Jm}

\maketitle

\section{Introduction}

Phases of many-body systems are often characterized in terms
of the long-distance
behavior of spatial correlation functions.\cite{sachdev_book,chaikin}
In particular, the decay of a correlation function to a finite value at
long distance is characteristic of an ordered phase, exponential decay
indicates a gapped or disordered phase, while power-law 
decay is a characteristic feature of critical systems. 
In quantum systems, 
entanglement-based measures can also be used to
characterize phases.\cite{HHHH,SzSzDiss}
In the past decade, much effort has been devoted to
developing and
understanding quantitative measures of
entanglement.\cite{luigi2008,wooters98,VidalEntMon,HorodeckiEntMeas}
Our goal here is 
to elucidate the relationship
between traditional correlation functions and such entanglement-based
measures of quantum correlations.

The full description of a quantum mechanical system 
can be given in a pure
state by the wavefunction, $|\Psi\rangle$, or in a mixed state 
by the density matrix, $\rho$.
For a bipartite system in a pure state, information on the
entanglement of the two parts of the system is encoded in the reduced
density matrix of one subsystem,
$\rho_A = \mbox{Tr}_B \, |\Psi\rangle\langle\Psi|$,
where we label the subsystem of interest by A and the other 
subsystem by B,
and $\mbox{Tr}_B$ means carrying out the trace over subsystem B.
The eigenvalue spectrum of $\rho_A$ completely characterizes the
entanglement between subsystems A and B.
A number of quantitative measures of entanglement can be extracted
from the eigenvalue spectrum.\cite{VidalEntMon,HorodeckiEntMeas}
The most commonly used measure is the von Neumann entropy
\cite{OhyaPetzQEntr,PetzQInfo}
\begin{equation}
  s_A = - \mbox{Tr} \rho_A \ln \rho_A \, ;
\label{eq:vonN_entropy}
\end{equation}
others include the more general R\'enyi entropy,\cite{renyi} the
concurrence,\cite{wooters98} and the Schmidt number.
Once the eigenvalues $w_{\alpha}$ of
$\rho_A$ are known, the von Neumann entropy can be calculated
using
$s_A=-\sum_{\alpha} w_{\alpha}\ln w_{\alpha}$.

The Hilbert space of a finite many-body quantum system
is formed by taking the tensor product of simple
building-block systems, which will be called ``sites''; here we
generally take the sites to be local units with ``finite Hilbert spaces''
 on a regular lattice.
In Eq.~(\ref{eq:vonN_entropy}), subsystem $A$ can be formed, in general,
from an arbitrary subset of the total set of sites.
While the usual practice is to take one, two, or more neighboring
sites, it can also be useful to form it from non-adjacent 
sites.~\cite{GabrielMurgHiesmayrMPS}

The number of sites included and where they are located on the
lattice can be tailored to obtain specific information on the
distribution of entanglement on the lattice, which can then be used to
characterize the physical nature of the quantum phase of the
system.
For example, the scaling behavior of the von Neumann entropy
of a contiguous block of sites with the number
of sites has been used to study the quantum phases of
one-dimensional systems.
For systems with local interactions, this ``block entropy'' diverges
logarithmically with block size for critical systems, but saturates
for gapped systems,\cite{vidallatorre03,calabrese04} and 
it has more complex behavior when 
non-local interactions are present.\cite{legeza2003b,barcza2010}

The entropy of particular subsystems such as 
the single-site, the nearest-neighbor two-site, and the block entropy
as a function of microscopic control parameters
can be used to detect and characterize many classes of quantum
phase transitions.
\cite{luigi2008,calabrese04,vidal,mfyang,legeza2006,laflorence,legeza2007,braganca2014}
However, since topological states cannot  be described by local order 
parameters, quantum phase transitions to such states must be detected by alternative
means.\cite{wen1990,kitaev2006,levin2006,eisert2010,hamma2008,pollman2012}
For example, for two-dimensional systems with topological order such as the
Heisenberg model on a kagome lattice, deviations of the block entropy
from area-law scaling can be used to 
detect topologically ordered phases.\cite{jiang2012,schollwoeckzz,whiteyy}
In addition, a gap in the entanglement spectrum 
can be used to characterize topological order.\cite{haldane_ES}

In the cases discussed above, the entanglement has been
 measured between two
subsystems of a bipartite partition of the total system.
Thus, it characterizes all correlations 
of quantum origin 
when the total system is in a pure state. 
While the dependence of these subsystem entropies on size and
placement on the lattice can yield some information on spatial
properties, they do not contain specific information on the 
entanglement between sites.

Taking into account the amount and structure of entanglement 
between different part of a large system is also very
important in developing numerical algorithms based on tensor product approximations.
\cite{white1992,schollwock2005,manmana2005,hallberg2006,schollwock2011,verstraetecirac04,murgverstraete07,murgverstraete08,verstraeteciracmurg08,vidal06,changlani09,marti10,murg-tree,chan2013-tree,murg2014}
Since the computational
cost of all of these methods is determined by the rank of the matrices
\cite{legeza2003a,legeza2003b} 
and tensors,\cite{hackbush} it is crucial
to obtain as much knowledge as possible about the entanglement
structure of the system under study and to incorporate this knowledge
in the matrix or tensor structure of the state used to simulate it.\cite{murg2014}

In the present paper, we will consider correlations between two sites.
Since these sites are embedded in a larger system, they are, in general,
in a mixed state, leading to a more complicated picture of the origin
of the correlations.
The correlations can be of classical or of quantum origin; moreover, there
are quantum correlations 
which are not due to entanglement.\cite{DornerVedralCorr}
A useful quantity to numerically characterize all kinds of
correlations between pairs of sites is the mutual information 
\begin{equation}
I_{ij} = s_i + s_j - s_{ij} \,,
\label{eq:mut}
\end{equation}
calculated between two generally placed sites, $i$ and $j$.
Here $s_i$ is the von Neumann entropy
defined in Eq.~(\ref{eq:vonN_entropy}), where now
a subsytem A is chosen to be the single site $i$, and $s_{ij}$ is the
entropy of the subsystem consisting of
sites $i$ and $j$.
The mutual information $I_{ij}$ describes the 
correlation between sites $i$ and $j$ embedded in a larger 
system.
When studied for all pairs $i$ and $j$, it yields a weighted graph of the
overall 
correlations of both classical and quantum origin in the lattice.
The mutual information defined in this way has been used 
previously to study correlation between
neighboring sites in spin and 
fermionic chains with local interactions \cite{legeza2005} and in quantum 
chemical problems.
\cite{rissler2006,barcza2010,boguslawski2012a,boguslawski2012b,boguslawski2013,boguslawski2014}
Here we will study its 
relationship to
conventional two-point correlation functions 
in quantum lattice models.

We will introduce transition operators that in a given basis of 
single-site states of a many-body system have a single non-vanishing 
matrix element.  
We will then show that the matrix elements of the two-site
density matrix needed to calculate the mutual information can be 
expressed in terms of expectation values of transition operators.
These matrix elements have the form of
{\it generalized} correlation functions and contain, by definition,
all two-site correlations.
Thus, there is no need for {\it a priori} knowledge 
about the nature of the ordering in the system, i.e., no need to
explicitly construct particular physical correlation functions.
The two-site mutual information in terms of the von Neumann entropy
consists of a weighted average
of generalized correlation functions and, in fact,
measures the strength of the overall correlation 
between sites $i$ and $j$. 
In the following, we will call this  
an ``entanglement bond'' if it is larger than a 
pre-defined threshold
value, chosen here to be $10^{-1}$.
The two-site mutual information, however, 
does not explicitly tell us what physical process is
responsible for the entanglement in the system.
The task, then, is to relate the generalized correlation functions to
physically motivated correlation functions such as spin-spin,
density-density, electron-hole, or pairing correlation functions.

Within the context of a discussion of the entropy area law for
classes of  classical and quantum lattice models with short-range
Hamiltonians, Wolf {\em et al.} \cite{wolf2008} 
derived
an inequality (their Eq.\ (5)) in which the
mutual information of two spatially separated regions is bounded from
below by the (appropriately normalized) square of any 
connected correlation function between local operators in the two
regions.
The implication is that the equality holds at large separations when
the most slowly decaying correlation function is inserted.
Our study of the two-site mutual
information finds that the equality does indeed apply for the most
slowly decaying correlation function, consistent with this bound.
In addition, for conformally invariant systems, Furukawa 
{\em et  al.}\cite{furukawa2009} studied the behavior of the mutual
information of two separated subsystems of general size and related 
its behavior to  the square of the
exponent of the leading algebraic decay of the two-point correlation
functions.
Our findings below for conformally invariant phases are consistent
with this result.

In summary, our goal is, on the one hand, to provide an
explicit scheme for 
efficiently contructing the two-site mutual information for the 
the density-matrix renormalization group (DMRG) 
and other matrix-product-state-(MPS)-based methods and, 
on the other, to investigate its
behavior for several conformally and non-conformally-invariant
models to make contact with the theoretical predictions in
Refs. \onlinecite{wolf2008} and \onlinecite{calabrese04}.

The paper is organized as follows.
In Sec.~\ref{sec:theory}, we describe  
the theoretical background.
In Sec.\ \ref{sec:spinhalf}, we demonstrate our approach on the
one-dimensional, spin-1/2, anisotropic Heisenberg model.
In Sec.~\ref{sec:su_n}, we present the spatial entanglement pattern
for the SU($n$) Hubbard model at various commensurate fillings
and, based on the generalized correlation functions, identify the
underlying relevant physical processes.
Finally, Sec.~\ref{sec:conclusion} contains our conclusions.

\section{The one- and two-site density matrix}
\label{sec:theory}

\subsection{One-site density matrix and the transition operators}
\label{sec:theory_den_mats}

\begin{figure}[htb]
\vskip -0.1cm
\centerline{
  \includegraphics[scale=0.4]{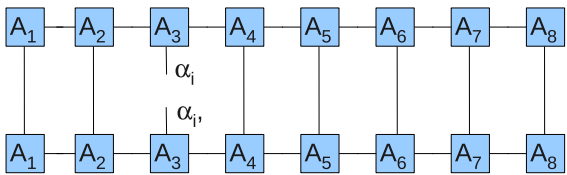}
}
\caption{Contraction of the MPS network to calculate the one-site reduced
  density matrix for a chain with $N=8$.}
\label{fig:mps_rho_i}
\end{figure}

The $N$-site wave function can be written in terms of the
single-site $q$-dimen\-si\-o\-nal basis as
\begin{equation}
\left|\Psi\right\rangle =\sum_{\alpha_1,\ldots,\alpha_N}
C_{\alpha_1,\ldots\alpha_N}\left|\alpha_1\rangle\ldots|\alpha_N\right\rangle \, ,
\label{eq:fulltensor}
\end{equation}
where the $\alpha_{j}$ denote single-site basis states and 
the set of coefficients $C_{\alpha_1,\ldots,\alpha_N}$ can be viewed as an
Nth-order tensor.
Matrix elements of the single-site density matrix
$\langle \alpha^\prime_i | \rho_i | \alpha_i \rangle $ 
are calculated 
by taking the trace of $|\Psi\rangle\langle\Psi|$ over all local bases
except for $\alpha_i$, the bases of sites $i$. 
The dimension of $C$ grows exponentially with system size $N$; thus,
such full tensor representations of the wave function are only possible
for small system sizes. 
Fortunately, the $N$th-order tensor $C$ can, in many cases, be efficiently
factorized into a product of matrices, 
$C_{\alpha_1,\ldots,\alpha_N}=A_{\alpha_1}\ldots A_{\alpha_N}$,
leading to a MPS representation of the wave function, where the
$A_{\alpha_i}$ are $M\times M$ matrices in general.\cite{verstraetecirac04}
For systems with open boundary conditions, 
$A_{\alpha_1}$ and $A_{\alpha_N}$ are row and column vectors, respectively. 
In the MPS representation, the calculation of
$\rho_{i}$ is straightforward as it
corresponds to the contraction of the
network over all states except those at site $i$.
A diagram depicting such a contraction for a finite, $N=8$-site system 
is shown in Fig.~\ref{fig:mps_rho_i}.
In the MPS-based two-site DMRG method, the system is divided into four
subsystems, and $|\Psi\rangle$ is formed from the tensor products of
the corresponding Hilbert spaces
(see Ref.~[\onlinecite{schollwock2005}]).
%OL 20150831
%\begin{equation}
%|\Psi\rangle=\sum_{\substack{\alpha_l,\alpha_{l+1},\\
%                     \alpha_{l+2},\alpha_r}}
%C_{\alpha_l,\alpha_{l+1},\alpha_{l+2},\alpha_r}
%\, |\phi^{(l)}_{\alpha_l}\rangle \, |\alpha_{l+1}\rangle\,
%|\alpha_{l+2}\rangle\, |\phi^{(r)}_{\alpha_r}\rangle\,.
%\label{eq:psi_dmrg}
%\end{equation}
For a given partitioning, expectation values can be calculated within
the multi-site block basis.

In general, the trace to obtain $\rho_i$, as carried out over the
MPS wave function 
can be decomposed into a sum of projection operators related to the
bases described by the free variables $\alpha_i$.
That is, the matrix representation of the one-site  reduced
density matrix $\langle \alpha^\prime_i | \rho_i | \alpha_i \rangle $ 
can be constructed from operators describing transitions
between the single-site basis $|\alpha_i\rangle$.

For a $q$-dimensional local Hilbert space, we can represent transitions
between the basis states using $q^2$ possible transition operators 
${\cal T}^{(m)}$ with $m=1\cdots q^2$.
The structure of the $q\times q$ operators ${\cal T}^{(m)}$
is simple: each operator contains only a single non-vanishing
matrix element 
with the value one at position $\alpha^\prime$, $\alpha$, where $|\alpha\rangle$ is
the initial state and $|\alpha^\prime\rangle$ is the final state; explicitly,
$({\cal T}^{(m)})_{\alpha^\prime,\alpha}=\delta_{\alpha+q[\alpha^\prime-1],m}$. 
It thus acts like a transition matrix between state $|\alpha\rangle$ and 
$|\alpha^\prime\rangle$, 
${\cal T}^{(m)}|\alpha\rangle=|\alpha^\prime\rangle$,
for
$\alpha = (m-1)({\rm mod } \;\, q)+1$ %initial state
and
$\alpha^\prime=\left \lfloor (m-1)/q \right \rfloor +1$, %final state
where $\left \lfloor x \right \rfloor$ denotes the floor function, the
integral part of $x$.
These operators can be extended to operate on the complete Hilbert
space consisting of $N$ local Hilbert spaces labeled by $i=1, \ldots, N$
as
\begin{equation}
{\cal T}_i^{(m)}=\bigotimes_{j=1}^{i-1} \openone \otimes 
{\cal T}^{(m)}  \otimes \bigotimes_{j=i+1}^{N} \openone \, ,
\end{equation}
where the operators on the right-hand side act on the appropriate
local single-site basis, and $\openone$ is the identity operator on
the local basis. 

For the many-electron wave function, 
the matrix elements of the one-site reduced density matrices
can be expressed in terms of expectation values of
${\cal T}_i^{(m)}$ operators acting on specific sites.
When the individual local states are completely
distinguished by abelian quantum numbers, the one-site density
matrix is diagonal and has the form
$
\langle \alpha^\prime_i | \rho_i | \alpha_i \rangle 
 = \delta_{\alpha_i,\alpha_i^\prime}
\langle {\cal T}_i^{(\alpha_i(q+1)-q)} \rangle$, 
where the expectation values are calculated using the $N$-site
many-body state.
Once the one-site reduced density matrix $\rho_i$ is constructed,
$s_i$ can be determined from its 
eigenvalues,
which are the selected expectation values themselves, hence
$s_i=-\sum_{l=1}^q\langle{\cal T}_i^{l(q+1)-q}\rangle\times
\ln \langle{\cal T}_i^{l(q+1)-q}\rangle$.

\subsection{Two-site density matrix and the generalized correlation functions}

We now consider the two-site reduced density matrix $\rho_{ij}$.
It can be calculated by
taking the trace of $|\Psi\rangle\langle\Psi|$ over all local bases
except for $\alpha_i$ and $\alpha_j$, the bases of sites $i$ and
$j$. 
In the MPS representation, the calculation of $\rho_{ij}$
corresponds to product of bra- and ket-MPS with two sites left uncontracted.
Its matrix elements 
$\langle \alpha_i^\prime,\beta_j^\prime | \, \rho_{ij}\,|\alpha_i,\beta_j\rangle$
mediate a transition from
state 
$|\alpha_i,\beta_j\rangle$
to state
$|\alpha_i^\prime,\beta_j^\prime\rangle$,
where $\alpha_i$ and 
$\alpha_i^\prime$ label the initial and final states,
respectively, on spatial site $i$ and $\beta_j$ and
$\beta_j^\prime$ the initial and final 
states, respectively, on spatial site $j$.
Thus, $\rho_{ij}$ can be calculated from expectation values of operator
products ${\cal T}^{(m)}_i{\cal T}^{(n)}_j$.
Here the two $q$-dimensional spaces for states defined on sites $i$
and $j$ are expressed as one $q^2$-dimensional space.
Since the operators ${\cal T}^{(m)}_i$ can be expressed in terms of
localized spin and fermion operators,
the two-site reduced density matrix
$\rho_{ij}$ can be built explicitly from the expectation values of
two-site operator products.
The two-site reduced density matrix $\rho_{ij}$ has non-zero matrix
elements only between two-site states possessing the same
quantum numbers 
because $\rho_{ij}$ does not change the quantum numbers of
the two sites (i.e., $\rho_{ij}$ is block-diagonal in the two-site
product basis), reducing the number of matrix elements that need be
calculated.
The symmetry properties of $\rho_{ij}$ further
reduce the number of expectation values to be determined. 

Once the two-site reduced density matrix $\rho_{ij}$ is constructed,
$s_{ij}$ can be determined from its 
eigenvalues $w_{ij,\alpha}$, 
and the mutual information can be calculated using Eq.~(\ref{eq:mut}).
An important feature of this method is that one can also
analyze the sources of entanglement encoded in $I_{ij}$ by studying
the behavior of the matrix elements of $\rho_{ij}$.
We term the expectation values of pairs of state-transition operators, 
$\langle {\cal T}_{i}^{(m)}\,{\cal T}_{j}^{(n)}\rangle=
\langle \Psi | {\cal T}_{i}^{(m)}\,{\cal T}_{j}^{(n)}| \Psi\rangle$,
{\em  generalized correlation functions} 
in order to distinguish them from  
conventional correlation functions, i.e., those based on physically
motivated operators such as local spin or density operators.
This is a generalization of the procedure introduced in the DMRG context for 
spin-1/2 fermion models in Refs.~\onlinecite{rissler2006,boguslawski2013}.
As we will see below, they can be used to identify the relevant
physical processes that lead to the generation of the entanglement.
This procedure will be demonstrated on the spin-1/2 Heisenberg
model and on the SU($n$) Hubbard model for $n\ge 2$.

It is important to take into account that 
when $\rho_{ij}$ is calculated, the system is decomposed into three
subsystems.
For a tripartite system, the wave function can be written
\begin{equation}
\left|\Psi\right\rangle =\sum_{\alpha_i,\alpha_j,\beta}
C_{\alpha_i,\alpha_j,\beta}|\alpha_i\rangle|\alpha_j\rangle|\beta\rangle \, ,
\label{eq:fulltensor_ije}
\end{equation}
where $\alpha_i$ and $\alpha_j$ label the bases of the sites $i$
and $j$, and $\beta$ labels the basis of the environment, which is 
composed of the remaining sites.
A given generalized correlation function describes the transition
from a particular initial state to a particular final state.
It can be expressed in  terms of the coefficients of the corresponding
wave function in the tripartite basis as 
\begin{widetext}
\begin{equation}
\langle {\cal T}_{i}^{(m)}\,{\cal T}_{j}^{(n)}\rangle 
= \sum_{\alpha^\prime_i,\alpha^\prime_j,\beta^\prime} \sum_{\alpha_i,\alpha_j,\beta} 
C_{\alpha^\prime_i,\alpha^\prime_j,\beta^\prime}^*
C_{\alpha_i,\alpha_j,\beta} \,
\langle\alpha^\prime_i|\langle\alpha^\prime_j|\langle\beta^\prime | 
  {\cal T}_{i}^{(m)}{\cal T}_{j}^{(n)} | \alpha_i\rangle|\alpha_j\rangle|\beta\rangle%_c RMN ???
= \sum_{\beta} C^*_{l(m),%\mcorr{r(m)}{
l(n),\beta} C_{%\mcorr{l(n)}{
r(m),r(n),\beta} \,, 
\end{equation}
\end{widetext}
where $r(m)=(m-1)({\rm mod } \;\, q)+1$ 
and $l(m)=\left \lfloor (m-1)/q \right \rfloor +1$. %final state
A given generalized correlation
function measures the 
amplitude in the wavefunction
within a particular environment.
For example, it can be used to
characterize the singlet valence bond between sites
with spin-1/2 degrees of freedom.

In general, $\langle {\cal T}_{i}^{(m)}\,{\cal T}_{j}^{(n)}\rangle$ 
contains both connected and disconnected contributions between
subsystems $i$ and $j$.
Therefore, it can, in general, scale to a finite 
value as the distance $l=|i-j|$ is increased. 
Note that this can occur even if the many-body state is not
characterized by long-range order, i.e., even if the physical
correlation function goes to zero for large $l$.
In order to circumvent this behavior, we generally study the connected
part of the generalized correlation functions, 
\begin{equation}
\langle {\cal T}_{i}^{(m)}\,{\cal T}_{j}^{(n)}\rangle_{\rm C} =  
\langle {\cal T}_{i}^{(m)}\,{\cal T}_{j}^{(n)}\rangle- 
\langle {\cal T}_{i}^{(m)}\rangle \langle{\cal T}_{j}^{(n)}\rangle \, , 
\end{equation}
where the disconnected part, given by the product of the expectation
values of the local transition operators, is subtracted out.
Note that in Eq.~(\ref{eq:mut}), the mutual information is
formulated in such a way that the disconnected parts of the general
correlation functions do not contribute. 
However, there is no way to express $I_{ij}$ directly as a function of
the connected parts of the generalized correlation functions.  

The spatial behavior of the correlation functions depends esentially on the 
type of intermediate states for which the matrix elements of
${\cal T}_{i}^{(m)}$ and ${\cal T}_{j}^{(n)}$ are finite
between these states and the ground state. Different  
${\cal T}_{i}^{(m)}$ operators may transfer the ground state to the
same intermediate state 
and, therefore, different generalized correlation functions may have
similar properties. 
Therefore, we can classify the generalized correlation functions
according to their type of decay. 
This behavior will also be investigated in detail in the
following sections.
Note that the decay properties
of two-site correlation functions can also be obtained from
the transfer operator used in the MPS 
framework.~\cite{Schuch-2013,Wall-2015} 

\subsection{Example: spin-1/2 case}
\label{subsec:spin-half}
As an example, we first consider a spin-1/2 model. 
The local Hilbert space is two-dimensional and is spanned by the
spin-up basis state $|\uparrow\rangle$ and the spin-down basis state
$|\downarrow\rangle$. 
The action of the four possible ${\cal T}^{(m)}$ 
(for $m=1\dots 4$) and their representations as standard
one-site operators are given as
\begin{align}
{\cal T}^{(1)}_i & =  |\downarrow\rangle\langle\downarrow| = -S^z_i+\frac{1}{2}\openone,\;
&{\cal T}^{(2)}_i &= |\downarrow\rangle\langle\uparrow|=S^{-}_i,\;\nonumber\\
{\cal T}^{(3)}_i &= |\uparrow\rangle\langle\downarrow|=S^{+}_i,\;
&{\cal T}^{(4)}_i &= |\uparrow\rangle\langle\uparrow| = S^z_i+\frac{1}{2}\openone.
\label{eq:op-s12}
\end{align}

Using these definitions,
we can express the physical correlation functions, $G_{ij}$,
directly in terms of the generalized correlation functions,
\begin{eqnarray}
G^{xx}_{ij} = \langle S^x_i \, S^x_j\rangle &=&  
\frac{1}{4} \left( \langle \mathcal{T}^{(2)}_{i}\mathcal{T}^{(3)}_{j}\rangle +\langle \mathcal{T}^{(3)}_{i}\mathcal{T}^{(2)}_{j}\rangle \right. \nonumber\\
&+&\left. \langle \mathcal{T}^{(2)}_{i}\mathcal{T}^{(2)}_{j}\rangle+\langle \mathcal{T}^{(3)}_{i}\mathcal{T}^{(3)}_{j}\rangle \right) \, ,\nonumber\\
G^{yy}_{ij} = \langle S^y_i \, S^y_j\rangle  &=&
\frac{1}{4}\left( \langle \mathcal{T}^{(2)}_{i}\mathcal{T}^{(3)}_{j}\rangle +\langle \mathcal{T}^{(3)}_{i}\mathcal{T}^{(2)}_{j}\rangle \right. \nonumber\\
&-&\left. \langle \mathcal{T}^{(2)}_{i}\mathcal{T}^{(2)}_{j}\rangle-\langle \mathcal{T}^{(3)}_{i}\mathcal{T}^{(3)}_{j}\rangle\right) \, ,\nonumber\\
G^{zz}_{ij} = \langle S^z_i \, S^z_j\rangle  &=& 
\frac{1}{4} \left( \langle \mathcal{T}^{(1)}_{i}\mathcal{T}^{(1)}_{j}\rangle +\langle \mathcal{T}^{(4)}_{i}\mathcal{T}^{(4)}_{j}\rangle \right. \nonumber\\
&-&\left. \langle \mathcal{T}^{(1)}_{i}\mathcal{T}^{(4)}_{j}\rangle-\langle \mathcal{T}^{(4)}_{i}\mathcal{T}^{(1)}_{j}\rangle \right).
\label{eq:G_spin_12}
\end{eqnarray}
For the most general spin Hamiltonian, all generalized correlation
functions might be different.
%RMN We do consider other models "in the rest of the paper"...
In the following,
we will consider the XXZ Heisenberg model (see Eq.~(\ref{eq:ham_j1j2}))
$\langle{T}^{(2)}_{i}\mathcal{T}^{(2)}_{j}\rangle=\langle \mathcal{T}^{(3)}_{i}\mathcal{T}^{(3)}_{j}\rangle=0$
and 
$\langle{T}^{(2)}_{i}\mathcal{T}^{(3)}_{j}\rangle=\langle \mathcal{T}^{(3)}_{i}\mathcal{T}^{(2)}_{j}\rangle$.
In addition, due to up-down symmetry in the region $-1<\Delta<1$,
$\langle{T}^{(1)}_{i}\mathcal{T}^{(1)}_{j}\rangle$, $\langle \mathcal{T}^{(1)}_{i}\mathcal{T}^{(4)}_{j}\rangle$,
$\langle{T}^{(4)}_{i}\mathcal{T}^{(1)}_{j}\rangle$, $\langle \mathcal{T}^{(4)}_{i}\mathcal{T}^{(4)}_{j}\rangle$
are not independent and they are related to $\langle S_i^z\rangle$ and $\langle S_j^z\rangle$.
On the other hand, as was mentioned earlier, the matrix elements of
$\rho_{ij}$ can be expressed  in terms of the same 
generalized correlation functions. The non-vanishing matrix elements are shown for the XXZ model in 
Table~\ref{tab:rho2-s12}, where, for easier readability, the notation
$m/n$ is used for 
$\langle{\cal T}^{(m)}_{i}{\cal T}^{(n)}_{j}\rangle$.
\begin{table}[!htb]
\centering
\begin{tabular}{c||c|cc|c}
\hline
\hline
$\rho_{ij}$ & {$|\!\downarrow\rangle_i$} {$|\!\downarrow\rangle_j\,$} & {$|\!\downarrow\rangle_i$} {$|\!\uparrow\rangle_j$} 
            & {$|\!\uparrow\rangle_i$} {$|\!\downarrow\rangle_j\,$} & {$|\!\uparrow\rangle_i$} {$|\!\uparrow\rangle_j$}\\
\hline
\hline
{$|\!\downarrow\rangle_i$} {$|\!\downarrow\rangle_j$} &  $1/1$ & & &\\
\hline
{$|\!\downarrow\rangle_i$} {$|\!\uparrow\rangle_j$}   &  & $1/4$ &   $2/3$ &  \\
{$|\!\uparrow\rangle_i$} {$|\!\downarrow\rangle_j$}   &  & $3/2$ &   $4/1$ &  \\
\hline
{$|\!\uparrow\rangle_i$} {$|\!\uparrow\rangle_j$}     &  &  &  & $4/4$\\
\hline
\hline
\end{tabular}
\caption{Block-diagonal form of the two-site reduced density matrix
  expressed in terms of generalized correlation functions. 
  Here $m/n$ denotes
  $\left\langle \Psi\right| {\cal T}_{i}^{(m)}\,{\cal T}_{j}^{(n)}
  \left|\Psi\right\rangle $. 
}
\label{tab:rho2-s12}
\end{table}

The eigenvalues of the two-site density matrix 
$\omega^{(\alpha)}$ ($\alpha=1,\ldots, q^2$) can 
easily be calculated.
Two of the eigenvalues are determined by the $1\times 1$ blocks given
by $\langle{\cal T}^{(1)}_i{\cal T}^{(1)}_{j}\rangle$ and 
$\langle{\cal T}^{(4)}_i{\cal T}^{(4)}_{j}\rangle$  
and can be expressed as 
\begin{eqnarray}
 \omega^{(1)}_{ij}&=&\frac{1}{4}-\frac{1}{2} \langle S^z_{i}
 \rangle-\frac{1}{2} \langle S^z_{j} \rangle+ \langle S^z_{i}S^z_{j}
 \rangle\, ,\\
 \omega^{(4)}_{ij}&=&\frac{1}{4}+\frac{1}{2} \langle S^z_{i}
 \rangle+\frac{1}{2} \langle S^z_{j} \rangle+ \langle S^z_{i}S^z_{j}
 \rangle \, .
\label{eq:eigvals-14}
\end{eqnarray}
Here we consider specifically the $S^z \equiv \sum_i S^z_i=0$ sector for which
$\langle S^z_{i} \rangle=0$; the expressions above can then be further simplified.
Similar calculation leads to 
\begin{equation}
 \omega_{ij}^{(2,3)}=1/4- \langle S^z_{i}S^z_{j} \rangle \pm \langle
 S^+_{i}S^-_{j} \rangle \, .
\label{eq:eigvals-23}
\end{equation}
The entropies $s_{ij}$, $s_i$, and $s_j$ as well as the mutual
information can be calculated
from the eigenvalue spectrum of the one- and two-site reduced density matrix.
For example, when 
sites $i$ and $j$ are uncorrelated, both
$\langle S^z_{i}S^z_{j} \rangle$ and $\langle S^+_{i}S^-_{j} \rangle$ vanish,
and $\omega_{ij}^{(\alpha)}=1/4$ for all $\alpha$. 
This means that the two-site subsystem is in a maximally mixed state,
and all $q^2$ states are equally probable. 
As a consequence, $I_{ij}=0$ since $s_{ij}=\ln 4$ and $s_i=s_j=\ln 2$.

For the ferromagnetic ground state, an example of an
ordered state, $\langle S^z_{i}S^z_{j} \rangle=1/4$,  
$\langle S^z_{i}\rangle=\langle S^z_{j}\rangle=1/2$,  
and $\langle S^+_{i}S^-_{j} \rangle=0$; thus, $\omega_{ij}^{(4)}=1$ and
$\omega_{ij}^{(\alpha)}=0$ for $\alpha=1,2,3$. 
Since the two-site subsystem is in a pure state, it is fully separable
from the rest of the system.
The correlation between sites $i$ and $j$, as given by the mutual
information $I_{ij}$, is also zero, since $s_i=s_j=0$.

The spatial behavior of the mutual information can be analyzed as 
a function of the distance $l=|i-j|$.  
In general, the eigenvalues 
depend on the distance $l$ and their decay to zero or a constant value 
can be analyzed. 
The decay rate for a given eigenvalue can depend on more
than one correlation function.
Therefore, in order to determine the exponent of the decay of the
eigenvalues in the $l\rightarrow\infty$ limit,
one has to analyze all the generalized correlation functions that
comprise a particular eigenvalue.

The asymptotic behavior of the mutual informaton can also be
determined in terms of the asymptotic behavior of the generalized
correlation functions.
When the expectation values of the generalized 
correlation functions become small for large distance $l=|i-j|$,
the von Neumann entropy for two sites, 
$s_{ij}=- \sum_\alpha \omega^{(\alpha)}_{ij} \ln \omega^{(\alpha)}_{ij}$ 
can be series-expanded using   
$\ln\left(1+x\right)\approx x-\frac{x^2}{2}$, when $x\ll 1$.  
In order to determine the leading terms in $s_{ij}$,
we keep the second-order terms in the series expansion and
obtain
to leading order,
\begin{equation}
 s_{ij} \approx \ln4 - 8 \langle S^z_{i}S^z_{j} \rangle^2 - 4 \langle S^+_{i}S^-_{j} \rangle^2
\label{eq:s2-series}
\end{equation}
for large $l=|i-j|$. 
Since $s_i$ is a constant, it follows from Eqs.~(\ref{eq:mut})
and (\ref{eq:s2-series}) that $I_{ij}$ decays
as the square of the slowest decaying correlation function. 

This analysis can be extended to higher spin values and to models with
more than one quantum number.
For example, for a spin-one model, the two-site density matrix 
contains 19 different generalized two-site correlation
functions.
For a spin-1/2 fermionic model, 
36 different two-site general
correlation functions have to be
calculated;\cite{rissler2006,barcza2010}
they are given explicitly in
Ref.~\onlinecite{boguslawski2013} but are also summarized in 
Sec.~\ref{sec:su_n}.
For fermions with larger numbers of flavors, this number increases
significantly. 
To indicate how this occurs, we summarize the number
of nonzero matrix elements in the reduced two-site density matrix,
i.e., the number of operator combinations to be calculated, in
Table~\ref{tab:corrfuns}
for the models studied in the present work. 
Note that the number of non-vanishing independent correlation functions 
to be calculated can be reduced by taking into account up-down or
left-right symmetries.
\begin{table}[!htb]
\centering
\begin{tabular}{|l|c|c|}
\hline
Model & $q$ & $N_{\rm ops}$ \\
\hline
S=1/2 spin & 2 & 6  \\
S=1   spin & 3 & 19 \\
2-flavor fermion & 4 & 36\\ 
3-flavor   fermion & 8 & 216 \\
4-flavor fermion &16 & 1296 \\
5-flavor   fermion &32 & 7776 \\
\hline
\end{tabular}
\caption{Number of nonzero matrix elements of the two-site reduced
  density matrix for $S$-spin  bosonic and $S$-flavor fermion systems
  with local Hilbert space dimension $q$.}  
\label{tab:corrfuns}
\end{table}

We note that the procedure outlined above can be extended to obtain 
the $n$-site reduced density matrix. 
However, the structure of correlations as well as that of the entanglement
becomes much more complicated if more than two sites are
involved,\cite{SzalayKokenyesiPartSep} and, accordingly, the
relevant measures of correlations and their interpretations are not
well understood.
Nevertheless, the correlation in the respective subsystems can be
characterized by generalizations of the two-site mutual information. 
For example, the three-site correlation  
can be defined as a Venn-diagram-based three-site
generalization \cite{CoverThomasInfoTheory} of the two-site mutual
information, 
\begin{eqnarray}
I_{ijk} = s_i + s_j + s_k%\nonumber\\
- s_{ij} - s_{ik} - s_{jk} 
+ s_{ijk}\, .
\end{eqnarray}
The three-site reduced density matrix $\rho_{ijk}$ can be expressed
in a straightforward way; Table~\ref{tab:rho_ijk} gives its definition 
for the spin-1/2 case.

\renewcommand{\ups}{{$\uparrow$}}
\renewcommand{\downs}{{$\downarrow$}}
\begin{table}[!htb]
\centering\scalebox{0.95}{
\begin{tabular}{c||c|ccc|ccc|c}
\hline
\hline
 & \downs \downs \downs & \downs \downs \ups\!  & \downs \ups \downs\!  & \ups \downs \downs\!  & \downs \ups \ups & \ups \downs \ups & \ups \ups \downs & \ups \ups \ups \\
\hline
\hline
\downs \downs \downs & 1/1/1 & &  &  &  &  &  & \\
\hline
\downs \downs \ups   & & 1/1/4 & 1/2/3 & 2/1/3 &  &  &  & \\
\downs \ups \downs   & & 1/3/2 & 1/4/1 & 2/3/1 &  &  &  & \\
\ups \downs \downs   & & 3/1/2 & 3/2/1 & 4/1/1 &  &  &  & \\
\hline
\downs \ups \ups     & & & & & 1/4/4 & 2/3/4 & 2/4/3 & \\
\ups \downs \ups     & & & & & 3/2/4 & 4/1/4 & 4/2/3 & \\
\ups \ups \downs     & & & & & 3/4/2 & 4/3/2 & 4/4/1 & \\
\hline
\ups \ups \ups       & & & & & & & & 4/4/4 \\
\hline
\hline
\end{tabular}
}
\caption{Block-diagonal form of the three-site density matrix
  expressed in terms of correlation functions.
  Here \downs\downs\downs denotes the state $|$\downs$\rangle_i|$\downs$\rangle_j|$\downs$\rangle_k$ 
and $m/n/l$ denotes %${\cal T}^{(m/n/l)}_{ijk}=   
$\left\langle \Psi\right| {\cal T}_{i}^{(m)}\,{\cal T}_{j}^{(n)}{\cal T}_{k}^{(l)} \left|\Psi\right\rangle$. 
}
\label{tab:rho_ijk}
\end{table}
%

%OL 20150831
%\subsection{Efficient calculation of the two-site density matrix}

\section{Anisotropic spin-1/2  Heisenberg chain}
\label{sec:spinhalf}

We consider first the anisotropic spin-1/2 Heisenberg chain  
where the numerical results can 
be compared to exactly known behavior.
The Hamilton has the form
\begin{eqnarray}
  H &= &\sum_{i} \left[ \frac{1}{2} (S^+_i S^-_{i+1} + S^-_i S^+_{i+1} ) +\Delta S^z_i S^z_{i+1} \right] %\nonumber
\label{eq:ham_j1j2}
\end{eqnarray}
where $S^+_i$ and $S^-_i$ are the spin raising and lowering operators,
respectively, $S^z_i$ is the $z$ component of the spin, and  
$\Delta$ parameterizes the anisotropy. 
The ground state of the system is critical for $-1 < \Delta \le 1$, 
and it has ferromagnetic and antiferromagnetic order for
$\Delta \leq -1$ and $\Delta > 1$, respectively.

In order to characterize the behavior, we consider the critical
  exponents $\nu_a$ of the correlation
functions $\langle S^a_i \, S^a_j\rangle\sim|i-j|^{-\nu_a}$ with $a\in\{x,y,z\}$.
The values of these critical exponents have a known dependence on
$\Delta$:\cite{heis_exponents}
\begin{equation}
  \nu_x=\nu_y=1/\nu_z=1-\arccos(\Delta)/ \pi \, .
\label{eq:heis_exps}
\end{equation}
As discussed in Sec.\ \ref{subsec:spin-half}, the generalized correlation
functions 
       $\langle{\cal T}_{i}^{(m)}\,{\cal T}_{j}^{(n)}\rangle$
can be expressed directly in terms
of the spin-spin correlation functions $\langle S^z_i \, S^z_j\rangle$
and $\langle S^+_i \, S^-_j\rangle$, so that the $\nu_a$ in
Eq.~(\ref{eq:heis_exps}) should directly determine the
behavior of $\langle{\cal T}_{i}^{(m)}\,{\cal T}_{j}^{(n)}\rangle$.

\begin{figure}[!htb]
\centerline{
  \includegraphics[scale=0.47]{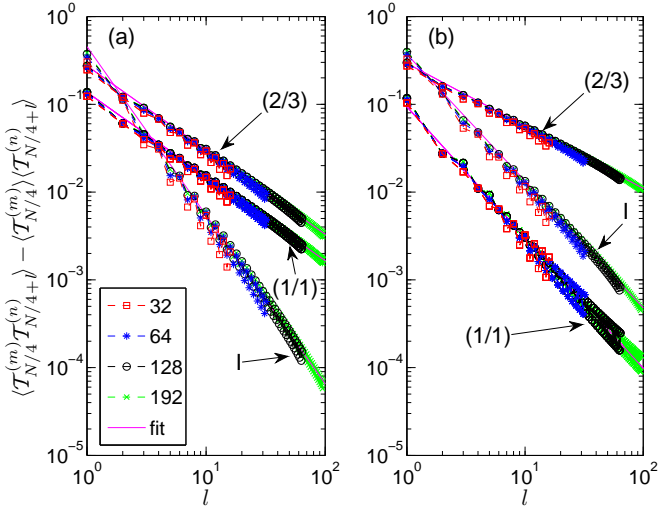}
}
\caption{(Color online) 
  Decay of the connected part 
  $\langle{\cal T}_{N/4}^{(m)}\,{\cal T}_{N/4+l}^{(n)}\rangle_{\rm C}$ of the 
  various two-site general correlation functions 
  used to construct the two-site density matrix 
  and the two-site mutual information  $I_{N/4,N/4+l}$ as a function
  of distance $l$ for various system sizes on a log-log scale for 
  (a) $\Delta=1$ and (b) $\Delta=0.5$.  
  The solid lines show linear fits used to obtain decay exponents  
  for $(m/n)=(2/3),(1/1)$, and $(4/4)$, and  
  for $I_{N/4,N/4+l}$.
}
\label{fig:I_scale_j2_0}
\end{figure}

The correlation functions depend essentially on 
the type of intermediate states for which the matrix elements of
${\cal T}_{i}^{(m)}$ and ${\cal T}_{j}^{(n)}$ are finite
between these states and the ground state. 
In conformal field theory, the two-site correlation functions fall into groups
characterized by the value of the exponent,
which we will denote by $\nu_{{\rm g}k}$, where $k$ labels the 
group number. 
%For example, for 
Although the XXZ Heisenberg chain is not conformally invariant, its
correlation functions decay with an oscillatory amplitude,
and the system has a tower structure in 
the excitation spectrum.
Eq.~(\ref{eq:heis_exps}) implies that there will be algebraic 
decay that falls into two groups in general, $\nu_x=\nu_y$ and 
$\nu_z=1/\nu_x$, which coalesce into one group for $\Delta=1$.
In Fig.~\ref{fig:I_scale_j2_0}(a), we have plotted 
the absolute value of the connected part 
$\langle{\cal T}_{i}^{(m)}\,{\cal T}_{j}^{(n)}\rangle_{\rm C}$ of the 
two independent generalized two-site correlation functions 
used to construct the two-site reduced density matrix (see
Table~\ref{tab:rho2-s12}) as well as
$I_{ij}$ itself.
Correlations measured in the bulk are plotted as a function of $j$ for  $i=N/4$ for $\Delta=1$ and
various system sizes on a log-log scale.
For the SU(2) symmetric case, $\Delta=1$, we find that 
both independent generalized
correlation functions scale to zero algebraically as
$l\rightarrow\infty$ with an exponent $\nu_{\rm g1}=0.95(3)$, 
 i.e.,
they do, in fact, form a single group.
The mutual information $I_{N/4,N/4+l}$ also 
decays to zero algebraically, $I_{N/4,N/4+l}\sim l^{\nu_I}$, but with a
different exponent, 
$\nu_{I}=1.97(3)\simeq 2\times\nu_{\rm g1}$. 
For the isotropic case 
$\langle S^+_i \, S^-_j\rangle=2\langle S_i^z \, S_j^z\rangle$,
the four eigenvalues derived from
Eqs.~(\ref{eq:eigvals-14},\ref{eq:eigvals-23}) are 
$\omega^{(1,2,4)}_{ij}=1/4+ \langle S^z_{i} \, S^z_{j} \rangle$,
$\omega^{(3)}_{ij}=1/4-3 \langle S^z_{i} \, S^z_{j} \rangle$,
and $s_{ij}=\ln 4 -24 \langle S^z_i \, S^z_j\rangle^2$,
so the exact value of $\nu_I=2$.

For $\Delta\ne 1$,
we summarize the content and behavior of the  
two groups of correlation functions, corresponding to longitudinal and
transverse correlations, 
in Table~\ref{tab:ops-group-s12} for $\Delta=0.5$. We obtain
$\nu_{\rm g1}=1.48(3)$ (exact value: $\nu_z=3/2$),
$\nu_{\rm g2}=0.69(4)$ (exact value $\nu_{x}=\nu_y=2/3$),
and $\nu_{I}=1.39(5)$ 
(exact value: $\nu_I = 2\times n_{\rm g1} = 4/3$).  
Thus, here we also 
find that the mutual information decays twice as fast as the
slowest decaying correlation function, as can be seen
explicitly in Fig.~\ref{fig:I_scale_j2_0}(b).
For a purely XX interaction, $\Delta=0$, we obtain
$\nu_{\rm g1}=0.52(5)$ for$\langle {\cal T}^{(2)}_1{\cal T}^{(3)}_{l}\rangle$ 
(exact value: $\nu_x=\nu_y=0.5$), while 
$\langle{\cal T}^{(1)}_1{\cal T}^{(1)}_l\rangle$, 
$\langle{\cal T}^{(1)}_1{\cal T}^{(4)}_l\rangle$, 
$\langle{\cal T}^{(4)}_1{\cal T}^{(1)}_l\rangle$, and 
$\langle{\cal T}^{(4)}_1{\cal T}^{(4)}_l\rangle$ decay with an exponent $\nu_{\rm g2}=2.15(8)$ 
(exact value: $\nu_z$=2). 
The mutual information decays with the exponent
$\nu_{I}=1.07(3) \simeq 2\times\nu_{g1}$.

Note that in order to obtain more accurate values for the critical
exponents, we would have to carry out an accurate finite-size scaling,
probably including non-leading effects such as logarithmic
corrections.\cite{hallberg1996}
When the U(1) symmetry is broken, for example, by switching on a magnetic 
field, 
$\langle{\cal T}^{(1)}{\cal T}^{(1)}\rangle\neq\langle{\cal T}^{(4)}{\cal T}^{(4)}\rangle$ and   
$\langle{\cal T}^{(1)}{\cal T}^{(4)}\rangle\neq\langle{\cal
  T}^{(4)}{\cal T}^{(1)}\rangle$.
In this case, the correlation functions break up completely into
different groups.
\begin{table}
\centering\scalebox{0.87}{
\begin{tabular}{|c|c|c|c|}
 \hline\multicolumn{4}{|c|}{Group-1 (algebraic), $\nu_{\rm g1}=0.69(4)$}\\
 \hline \hline
 Operator combination   & Second quantized form               & $\Delta S$ & $G$\\
 denoted as ($m/n$)     & ${\cal T}^{(m)}_i {\cal T}^{(n)}_j$ &                   & \\
 \hline
 (2/3) &  $S^-_i  S^+_j$ & $-1$ & $G^{xx}, G^{yy}$\\
 \hline
 (3/2) &  $S^+_i  S^-_j$ &  $1$ & $G^{xx}, G^{yy}$\\
 \hline
 \hline\multicolumn{4}{|c|}{Group-2 (algebraic), $\nu_{\rm g2}=-1.48(3)$}\\
 \hline \hline
 (1/1) &  $(-S^z+\frac{1}{2}\openone)_i (-S^z+\frac{1}{2}\openone)_j$ & 0 & $G^{zz}$\\
 \hline
 (1/4) &  $(-S^z+\frac{1}{2}\openone)_i  (S^z+\frac{1}{2}\openone)_j$ & 0 & $G^{zz}$\\
 \hline
 (4/1) &  $(S^z+\frac{1}{2}\openone)_i   (-S^z+\frac{1}{2}\openone)_j$ & 0 & $G^{zz}$\\
 \hline
 (4/4) &  $(S^z+\frac{1}{2}\openone)_i   (S^z+\frac{1}{2}\openone)_j$ & 0 & $G^{zz}$\\
 \hline
\multicolumn{4}{|c|}{Mutual information (algebraic), $\nu_I=-1.39(5)$}\\
 \hline
\end{tabular}
 }
\caption{Characterization of the two-site correlation functions based
  on their decay type and the value of their exponent for the spin-1/2
  anisotropic Heisenberg chain at $\Delta=0.5$ with $N=128$ sites.
  Here $\Delta S$ shows how ${\cal T}^{(m)}_i$ changes the spin
  quantum number.
  The corresponding conventional correlation functions
  as given in
  Eq.~(\ref{eq:G_spin_12}) are indicated by $G$.
  Operator combinations in group 1 correspond to transverse
  components and in group 2 to the longitudinal component of the spin correlations.
  The exact critical exponents are $\nu_x=\nu_y=2/3$ and 
  $\nu_z=3/2$.}
\label{tab:ops-group-s12}
\end{table}
In higher spin sectors, we find the same relationship, but   
additional oscillations appear in the decay of
the correlation functions, and 
$\langle S^z_i S^z_j\rangle$  
scales to a finite value  
due to the finite value of $S^z_{\rm tot} = \sum_i S_i^z$.

\section{SU$(n)$ Hubbard model}
\label{sec:su_n}

In this section, we apply our method to the SU$(n)$
symmetric Hubbard model for $n=2,3,4$, and $5$ for commensurate
fillings $f=p/q$, where $p$ and $q$ are relatively prime. 
The SU($n$) Hubbard model has the Hamiltonian
\begin{eqnarray} 
    {\mathcal H} & = & - t\sum_{i=1}^N\sum_{\sigma=1}^n (c_{i,\sigma}^\dagger
     c_{i+1,\sigma}^{\phantom \dagger} + c_{i+1,\sigma}^\dagger
     c_{i,\sigma}^{\phantom\dagger}) \nonumber \\
     &+& 
      \frac{U}{2}\sum_{i=1}^N \sum_{\substack{\sigma,\sigma'=1 \\ \sigma \neq
     \sigma'}}^n n_{i, \sigma}n_{i,\sigma'} \, ,
\label{eq:ham-sun}
\end{eqnarray} 
where $N$ is the number of sites in the chain. The operator
$c_{i,\sigma}^\dagger$ ($c_{i,\sigma}^{\phantom \dagger}$) creates
(annihilates) an electron at site $i$ with spin $\sigma$, where the spin index
is allowed to take on $n$ different values. 
Here $n_{i,\sigma}$ denotes the particle-number
operator, $t$ the hopping integral between nearest-neighbor sites, and $U$
the strength of the on-site Coulomb repulsion.  
For consistency, the spin (or, for $n>2$, flavor) index
of the operators will be indicated by a subscript index running from
$1$ to $n$, even for the SU(2) case. 
In what follows, we will take $t$ as the unit of energy.

By studying the length-dependence of the entropy of finite blocks 
of long chains, some of us showed in a previous work \cite{buchta_sun} that 
the system has drastically different behavior depending
on whether $q>n$, $q=n$, or $q<n$. 
In addition, by taking the Fourier transform of the oscillatory behavior
of the block entropy,\cite{legeza2007}
we were able to determine the position of soft modes in the excitation 
spectrum when the model is critical and the spatial inhomogeneity
of the ground state when the system is gapped.\cite{szirmai-2008}

When $q>n$, the umklapp processes are irrelevant, and the model
is equivalent to an $n$-component Luttinger liquid with central charge
$c=n$. 
When $q=n$, the charge and spin modes are decoupled, and the umklapp
processes open a charge gap for finite $U_{\rm c}>0$, while the spin
modes remain gapless and the central charge $c=n-1$. 
The value of $U_{\rm c}$ is still a subject of debate,
\cite{assaraf,buchta_sun,manmana-2011}
but it is known that the translational symmetry is not broken
in the ground state for any $n$.
 
\begin{figure}[!htb]
\centerline{
  \includegraphics[scale=0.6]{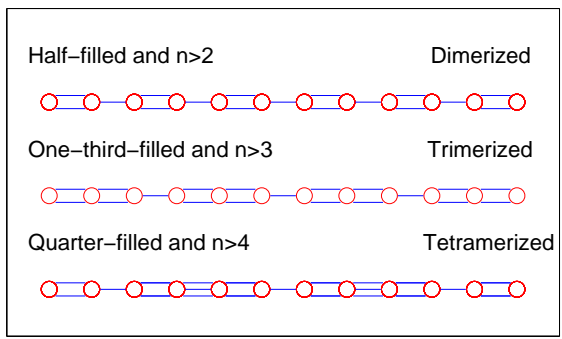}
}
\caption{(Color online) Depiction of the spatial inhomogeneity
  of the ground state of
  the SU($n$) Hubbard model in
  the thermodynamic limit for various fillings and $q<n$. 
  The strength of the bond is 
  indicated by the number of lines connecting nearest-neighbor
  sites.} 
\label{fig:su5_gstopology}
\end{figure}

On the other hand, when $q<n$, the charge and spin modes are coupled, 
the umklapp processes open gaps in all excitation branches, and a
spatially non-uniform ground state develops. 
Bond-ordered dimerized, trimerized, or tetramerized phases
can be found as the filling is varied.
These known results are summarized in Table~\ref{tab:analitikus},
and a schematic plot of the spatial inhomogeneity of the ground state 
determined using dimerization entropy is shown   
in Fig.~\ref{fig:su5_gstopology} for various fillings and
$n$ values.
\begin{table}[!htb]
\centering
\begin{tabular}{@{}ccccccccc@{}}  \hline \hline
& \phantom{+} & $n$ & \phantom{+}& $c$ & \phantom{+} & phase & \phantom{+} & $k^*$ \\ 
\hline 
$q=n$ & & any $n$ & & $n-1$ & & C0S$(n-1)$ & & $2\pi p/n$ \\ 
$q<n$ & & $n\neq 2$ & & -- & & C0S0 & & $2\pi p/q$ \\ 
$q>n$ & & any $n$ & & $n$ & & C1S$(n-1)$ & & $2\pi p/q$\\ 
\hline \hline
\end{tabular}
\caption{ 
  Central charge $c$ and type of phase, as
  characterized by the number of soft modes in the charge and spin
  sectors (CxSy), for the
  $p/q$-filled SU($n$) Hubbard model.}
\label{tab:analitikus}
\end{table}

This picture was obtained by calculating the spatial variation of the two-site
entropies of neighboring sites along the chain. In this paper we will
get a more quantitative description of the 
dimerized, trimerized, and quadrimerized phases
by calculating the spatial entanglement maps of the
two-site mutual information. Extracting such spatial entanglement
information and identifying the 
underlying physical processes will be  
the subject of the rest of this section. 

\subsection{The half-filled SU(2) Hubbard model}

The SU(2) Hubbard model is a free fermion model for $U=0$
and is thus trivially conformally invariant.
Therefore, the results discussed in the previous section
hold in this case.  
For finite $U$ and away from
half filling, the model is also conformally invariant, whereas  
for finite $U$ at half filling, the model is not conformally
invariant.
Our procedure, however, also applies
to non-conformally invariant models, as will be shown below.

For convenience, we use the standard notation $\uparrow$ and $\downarrow$
for the $\sigma=\{1,2\}$ spin states.
For a spin-1/2 fermionic model, single-electron basis states can be empty,
occupied with a single spin-down or spin-up electron,
or doubly occupied. 
These states we denote as $|$\zero$\!\rangle$, $|\!\downarrow\rangle$,
$|\!\uparrow\rangle$,
and $|\!\uparrow\downarrow\rangle$, respectively. 
Since the local basis is four-dimensional,
16 possible transition operators ${\cal T}_i^{(m)}$, as  
displayed  in
Table.~\ref{tab:primitiveop-f12}, arise.
They can be written explicitly in terms of local fermion
creation $c_{i\sigma}^\dagger$, annihilation $c_{i\sigma}$ and density $n_{i\sigma}$
operators as
\begin{align}
{\cal T}^{(1)}  &=(\openone-n_{\uparrow})(\openone-n_{\downarrow}),\;\;\nonumber
&{\cal T}^{(2)}  &= (\openone-n_{\uparrow}) c_{\downarrow},\;\;\\
{\cal T}^{(3)}  &= c_{\uparrow}(\openone-n_{\downarrow}),\;\;\nonumber
&{\cal T}^{(4)}   &= -c_{\uparrow}c_{\downarrow},\;\;\\
{\cal T}^{(5)}  &= (\openone-n_{\uparrow})c_{\downarrow}^{\dagger}   ,\;\;\nonumber
&{\cal T}^{(6)}   &= (\openone-n_{\uparrow})n_{\downarrow},\;\;\\
{\cal T}^{(7)}  &= - c_{\uparrow}c_{\downarrow}^{\dagger},\;\;\nonumber
&{\cal T}^{(8)}   &= c_{\uparrow}n_{\downarrow},\;\;\\
{\cal T}^{(9)}  &= c_{\uparrow}^{\dagger}(\openone-n_{\downarrow}),\;\;\nonumber
&{\cal T}^{(10)}  &= c_{\uparrow}^{\dagger}c_{\downarrow},\;\;\\
{\cal T}^{(11)} &= n_{\uparrow}(\openone-n_{\downarrow}),\;\;\nonumber
&{\cal T}^{(12)}  &= -n_{\uparrow}c_{\downarrow},\;\;\\
{\cal T}^{(13)} &= c_{\uparrow}^{\dagger}c_{\downarrow}^{\dagger},\;\;\nonumber
&{\cal T}^{(14)}  &= c_{\uparrow}^{\dagger}n_{\downarrow},\;\;\\
{\cal T}^{(15)} &= -n_{\uparrow}c_{\downarrow}^{\dagger},\;\;
&{\cal T}^{(16)}  &= n_{\uparrow}n_{\downarrow}.
\label{eq:fermion-ops-1}
\end{align}
Recalling that in this case 36 independent correlation functions
can be constructed (see Table~\ref{tab:corrfuns}), 
we will now relate them to $\rho_{ij}$.
Following the procedure outlined in Sec.\ \ref{sec:theory_den_mats},
the non-vanishing matrix elements of the two-site density matrix 
$\rho_{ij}$ are given in Table \ref{tab:fermion-ops-2}. 
Note that the two-site density matrix is block-diagonal in the
particle number $N_c$ and in the $z$ component of the spin $S_z$.
The block-diagonal structure is evident, and the values of $m$ and $n$
appropriate for each matrix element are displayed.
\begin{table}[!htb]
\centering
\begin{tabular}{c||c|c|c|c}
\hline
\hline
 & {$|$\zero$\rangle_i$} & {$|$\down$\rangle_i$} & {$|$\up$\rangle_i$} & {$|$\double$\rangle_i$} \\
\hline
\hline
{$|$\zero$\rangle_i$}  & ${\cal T}^{(1)}_i$ & ${\cal T}^{(2)}_i$ &  ${\cal T}^{(3)}_i$ & ${\cal T}^{(4)}_i$ \\
\hline
{$|$\down$\rangle_i$} & ${\cal T}^{(5)}_i$ & ${\cal T}^{(6)}_i$ & ${\cal T}^{(7)}_i$ & ${\cal T}^{(8)}_i$\\ 
\hline
{$|$\up$\rangle_i$}  & ${\cal T}^{(9)}_i$ & ${\cal T}^{(10)}_i$ &  ${\cal T}^{(11)}_i$ & ${\cal T}^{(12)}_i$ \\
\hline
{$|$\double$\rangle_i$} & ${\cal T}^{(13)}_i$ & ${\cal T}^{(14)}_i$ & ${\cal T}^{(15)}_i$ & ${\cal T}^{(16)}_i$\\ 
\hline
\hline
\end{tabular}
\caption{Single-site operators describing transitions
  between single-site basis states for a $S=1/2$ spin system.}
\label{tab:primitiveop-f12}
\end{table}
\begin{table*}[htb]%[b!]
\centerline{
 \includegraphics{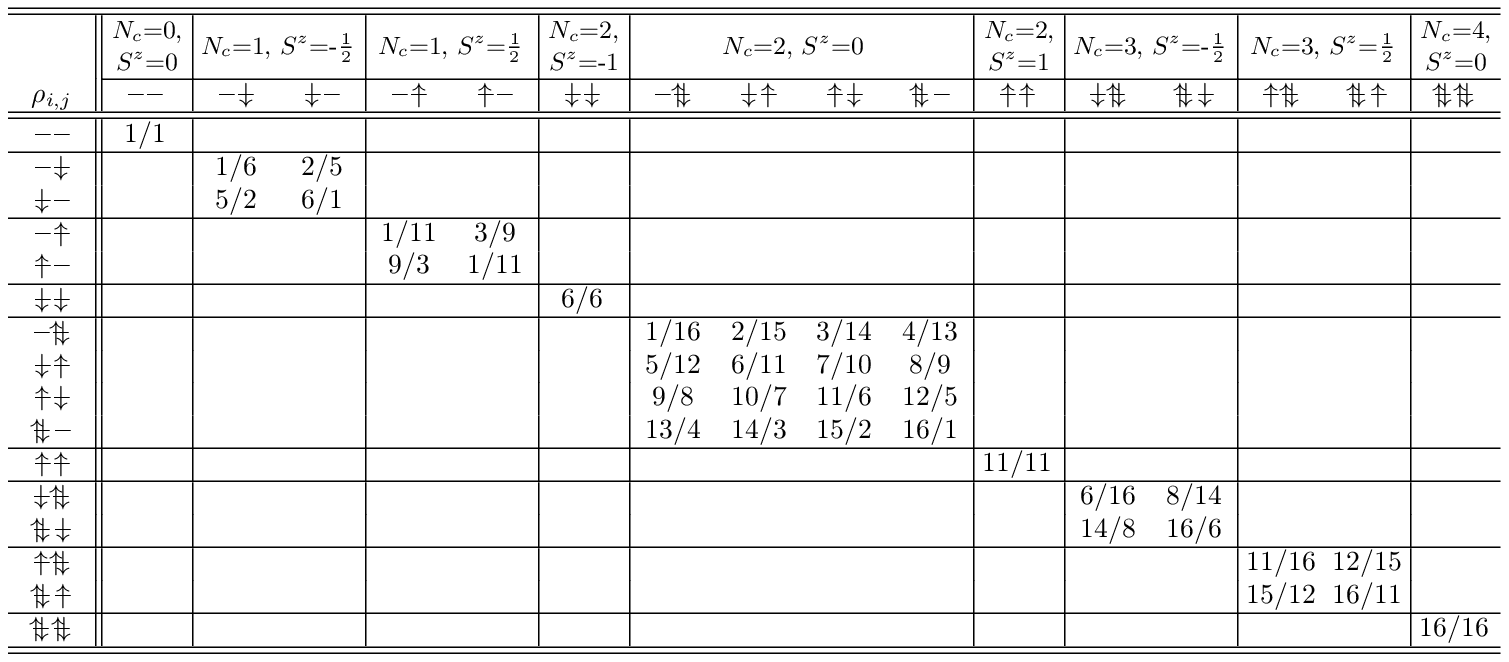}
}
\caption{The two-site reduced density matrix $\rho_{ij}$
  for SU(2) fermions expressed in terms of single-site
  operators, ${\cal T}^{(m)}_i$ with $m=1\cdots 16$. 
  For easier readability only the upper indices of the operators are
  shown; thus, $m/n$ denotes   
  $\langle {\cal T}^{(m)}_{i}{\cal T}^{(n)}_{j}\rangle $.
  Here $N_c$ and $S^z$ are the particle-number and $z$
    spin component quantum numbers of the two sites.
}
\label{tab:fermion-ops-2}
\end{table*}

We analyze the decay properties of the various generalized
correlation functions that appear in the operator decomposition  
of the reduced two-site density matrix, 
as given in Eq.~(\ref{eq:fermion-ops-1}) and 
Table~\ref{tab:fermion-ops-2},
and the two-site mutual information. 
\begin{table}[!t]
\centering
\begin{tabular}{|c|c|c|}
 \hline\multicolumn{3}{|c|}{Half-filled case}\\ 
 \hline \hline
$G$ & $U=0$ & $U>0$ \\
 \hline 
$G_{c^\dagger_\sigma c_\sigma}=\langle c^\dagger_\sigma \,  c_\sigma\rangle$ & $\nu=1$ & -- \\
$G_{n n}=\langle n \,  n \rangle$                                            & $\nu=2$ & 1 \\
$G^{zz}=\langle S^z \,  S^z \rangle$                            & $\nu=2$ & 1 \\
$G^{+-}=\langle S^+ \,  S^- \rangle$                         & $\nu=2$ & 1 \\
$G^{(0)}_{p}=\langle c^\dagger_1 c^\dagger_2 \,  c_2 c_1 \rangle$            & $\nu=2$ & -- \\
 \hline 
 \hline\multicolumn{3}{|c|}{Non half-filled case}\\ 
 \hline \hline
$G$ & $U=0$ & $U\gg W$ \\
 \hline 
$G_{c^\dagger_\sigma c_\sigma}=\langle c^\dagger_\sigma \, c_\sigma\rangle$ & $\nu=1$ & 9/8 \\
$G_{n n}=\langle n \, n \rangle$                                            & $\nu=2$ & 3/2 \\
$G^{zz}=\langle S^z \, S^z \rangle$                            & $\nu=2$ & 3/2 \\
$G^{+-}=\langle S^+ \, S^- \rangle$                         & $\nu=2$ & 3/2 \\
$G^{(0)}_{p}=\langle c^\dagger_1 c^\dagger_2 \, c_2 c_1 \rangle$            & $\nu=2$ & 5/2 \\
 \hline 
\end{tabular}
\caption{Conventional correlation functions and corresponding
  exact values of the critical exponents, where 
$n$ = $n_1+n_2, S^z=(n_1-n_2)/2$ and $S^+=c^\dagger_\sigma c_{\sigma^\prime}$ with $\sigma\in\{1,2\}\equiv\{\downarrow,\uparrow\}$
and $W = 4t$ is the bandwidth.}
\label{tab:su2-corr}
\end{table}
The exact critical exponents of conventional correlation functions, which have the form 
$G_{i,i+l}=\langle A_i \, A_{i+l}^\dagger\rangle\sim l^{-\nu_G}$, where
$A$ is a single-site spin, density, fermion-creation, or
pair-creation operator (see Table~\ref{tab:su2-corr}), are known from
the effective conformal field theory and are related to the behavior
of finite-size 
corrections to the ground state.\cite{korepin,woynarowich,korepin-ref7,schulz}
They are determined by the operator content of $A$, namely, by
the number 
of creation and annihilation operators for up- or
down-spin particles  
on the two branches (left and right) of the dispersion relation.
The exponents depend in a relatively simple manner on four parameters, i.e., 
on how the operator $A$ changes the number of electrons,
$\Delta N_c$,  
the spin, $\Delta S^z$, and on the value of charge current $J_c$ and
spin current $J_s$. 
In general, the exponents depend on $U$,
and are known exactly in the limiting cases. They are summarized in
Table~\ref{tab:su2-corr} for both the half-filled and the non-half-filled
systems.
At half filling, a metal-insulator transition takes place at $U=0$, 
leading to different behavior for $U=0$ and $U>0$.
For the non-half-filled case, the prefactor of the leading term, which
determines the critical exponent of the correlation function, 
may vanish in the $U\rightarrow\infty$ limit.\cite{korepin}
Therefore, 
the $U=\infty$ case must be treated separately. 
For example, for finite but large $U$,
$\nu_{G_{nn}}=3/2$, whereas $\nu_{G_{nn}}=2$ for $U=\infty$.  

We can express the conventional correlation functions as 
linear combinations of the generalized correlation functions.
For example, 
using the definitions of single-site transition operators as given in  
Eq.~(\ref{eq:fermion-ops-1}) and noting that 
\begin{eqnarray}
c^\dagger_\downarrow &=&c_\downarrow(\openone-n_\uparrow)+c_\downarrow n_\uparrow={\cal T}^{(5)}-{\cal  T}^{(15)}, \\ 
c^\dagger_\uparrow &=&c_\uparrow(\openone-n_\downarrow)+c_\uparrow n_\downarrow={\cal T}^{(9)}+{\cal  T}^{(14)}, \\  
c_\downarrow &=&c_\downarrow(\openone-n_\uparrow)+c_\downarrow n_\uparrow={\cal T}^{(2)}-{\cal  T}^{(12)}, \\ 
c_\uparrow &=&c_\uparrow(\openone-n_\downarrow)+c_\uparrow n_\downarrow={\cal T}^{(3)}+{\cal  T}^{(8)}, \\   
n_\downarrow &=&n_\downarrow(\openone-n_\uparrow)+n_\downarrow n_\uparrow={\cal T}^{(6)}+{\cal  T}^{(16)}, \\   
n_\uparrow &=&(\openone-n_\downarrow)n_\uparrow+n_\downarrow n_\uparrow={\cal T}^{(11)}+{\cal T}^{(16)}, \\ 
S^z   &=& \frac{1}{2}(n_\uparrow-n_\downarrow)=\frac{1}{2}\left(n_\uparrow(\openone-n_\downarrow)-n_\downarrow(\openone-n_\uparrow)\right)\nonumber\\
      &=& \frac{1}{2}({\cal T}^{(11)}-{\cal T}^{(6)})\, ,
\end{eqnarray}
the conventional correlation functions can be written as
\begin{eqnarray}
G_{nn}    
          &=&           \langle {\mathcal T}_{i}^{(6)} \, {\mathcal T}_{j}^{(6)} \rangle + 
          				\langle {\mathcal T}_{i}^{(6)}\, {\mathcal T}_{j}^{(11)} \rangle + 
                        2 \langle {\mathcal T}_{i}^{(6)}\,  {\mathcal T}_{j}^{(16)} \rangle  \nonumber \\
          &+&            \langle {\mathcal T}_{i}^{(11)}\,  {\mathcal T}_{j}^{(6)} \rangle + 
          				\langle {\mathcal T}_{i}^{(11)}\, {\mathcal T}_{j}^{(11)} \rangle + 
                         2 \langle {\mathcal T}_{i}^{(11)}\,  {\mathcal T}_{j}^{(16)} \rangle  \nonumber\\
          &+&             2 \langle {\mathcal T}_{i}^{(16)}\,  {\mathcal T}_{j}^{(6)} \rangle
          			 + 2 \langle {\mathcal T}_{i}^{(16)}\, {\mathcal T}_{j}^{(11)} \rangle + 
                         4 \langle {\mathcal T}_{i}^{(16)}\,  {\mathcal
                           T}_{j}^{(16)} \rangle \, ,\nonumber\\
\label{eq:G_nn}
G^{zz} &=&  \frac{1}{4}(\langle{\cal T}_{i}^{(11)} \, {\cal T}_{j}^{(11)}\rangle+
			\langle{\cal T}_{i}^{(6)} \, {\cal T}_{j}^{(6)}\rangle)\nonumber\\
      &-&  
           \frac{1}{4}(\langle{\cal T}_{i}^{(6)} \, {\cal T}_{j}^{(11)}\rangle+
           \langle{\cal T}_{i}^{(11)} \, {\cal T}_{j}^{(6)}\rangle) \, , \\
\label{eq:G_z}
G^{(+-)}&=&\langle {\cal T}_{i}^{(7)} \, {\cal T}_{j}^{(10)}\rangle\, ,\\
\label{eq:G_xy}
G_{c^\dagger_\downarrow c_\downarrow^{\phantom\dagger}}
 &=&  \langle{\cal T}_{i}^{(5)} \, {\cal T}_{j}^{(2)}\rangle-\langle{\cal T}_{i}^{(5)} \, {\cal T}_{j}^{(12)}\rangle\nonumber\\
 &-&  \langle{\cal T}_{i}^{(15)} \, {\cal T}_{j}^{(2)}\rangle+\langle{\cal T}_{i}^{(15)} \, {\cal T}_{j}^{(12)}\rangle\, , \\
\label{eq:G_cc}
G_{c^\dagger_\uparrow c_\uparrow^{\phantom\dagger}}
 &=&  \langle{\cal T}_{i}^{(9)} \, {\cal T}_{j}^{(3)}\rangle+\langle{\cal T}_{i}^{(9)} \, {\cal T}_{j}^{(8)}\rangle\nonumber\\
 &+&  \langle{\cal T}_{i}^{(14)} \, {\cal T}_{j}^{(3)}\rangle+\langle{\cal T}_{i}^{(14)} \, {\cal T}_{j}^{(8)}\rangle\, , \\
\label{eq:G_c2c2}
G^{(0)}_{p}&=&
\langle {\cal T}_{i}^{(13)} \, {\cal T}_{j}^{(4)}\rangle\, . %\\
\label{eq:G_1}
\end{eqnarray}
For $U=0$, all generalized correlation functions and thus
the mutual information decay algebraically due to soft modes at 
$k=0$ and $k=2k_{\rm F}$ in both the spin and charge sectors.
These soft modes lead to 
a $2k_{\rm F}$ oscillation in the leading terms. 
Grouping the operators according to their decay type and 
increasing value of the
exponent (see Table~\ref{tab:ops-group-su2u0}),
we identify three groups.  
%HERE appendix
%
\begin{table}
\scalebox{0.79}{
\begin{tabular}{|c|c|c|c|c|c|}
 \hline
($m/n$)  & Second quantized form & $\Delta{N_c}$ & $\Delta{S^z}$ & $G$\\
       & $A_i B_j$            &               &               &     \\
%&     \\
 \hline
\multicolumn{5}{|c|}{Group-1(algebraic)  $\nu_{g1}=0.98(4)$}\\
 \hline
 \hline
(12/15) &  $[n_\uparrow c_\downarrow]_i\; [n_\uparrow c^\dagger_\downarrow]_j$ &-1&0.5 & $G_{c^{\phantom\dagger}_\downarrow c^{\dagger}_\downarrow}$\\
 \hline
(2/15) &  $[(\openone-n_\uparrow)c_\downarrow]_i\; [-n_\uparrow c^\dagger_\downarrow]_j$ &-1&0.5&$G_{c^{\phantom\dagger}_\downarrow c^{\dagger}_\downarrow}$\\
 \hline
(2/5) &  $[(\openone-n_\uparrow)c_\downarrow]_i\; [(\openone-n_\uparrow)c^\dagger_\downarrow]_j$ &-1&0.5&$G_{c^{\phantom\dagger}_\downarrow c^{\dagger}_\downarrow}$\\
 \hline
(3/14) &  $[c_\uparrow(\openone-n_\downarrow)]_i\; [c^\dagger_\uparrow n_\downarrow]_j$ &-1&-0.5& $G_{c^{\phantom\dagger}_\uparrow c^{\dagger}_\uparrow}$\\
 \hline
(3/9) &  $[c_\uparrow (\openone-n_\downarrow)]_i\; [c^\dagger_\uparrow (\openone-n_\downarrow)]_j$ &-1&-0.5 & $G_{c^{\phantom\dagger}_\uparrow c^{\dagger}_\uparrow}$\\
 \hline
(5/12) &  $[(\openone- n_\uparrow) c^\dagger_\downarrow]_i\; [-n_\uparrow c_\downarrow]_j$ &1&-0.5 & $G_{c^\dagger_\downarrow c^{\phantom\dagger}_\downarrow}$\\
 \hline
(8/14) &  $[c_\uparrow n_\downarrow]_i\; [c^\dagger_\uparrow n_\downarrow]_j$ &-1&-0.5 & $G_{c^{\phantom\dagger}_\uparrow c^{\dagger}_\uparrow}$\\
 \hline
(8/9) &  $[c_\uparrow n_\downarrow]_i\; [c^\dagger_\uparrow (\openone-n_\downarrow)]_j$ &-1&-0.5 & $G_{c^{\phantom\dagger}_\uparrow c^{\dagger}_\uparrow}$\\
 \hline
\multicolumn{5}{|c|}{Group-2(algebraic)  $\nu_{g2}=1.92(5)$}\\
 \hline
 \hline
(11/11) &  $[n_\uparrow (\openone-n_\downarrow)]_i\; [n_\uparrow (\openone-n_\downarrow)]_j$ &0&0& $G_{nn}$, $G^{z}_{\sigma\sigma}$\\
 \hline
(16/16) &  $[ n_\uparrow n_\downarrow]_i\; [n_\uparrow n_\downarrow]_j$ &0&0 & $G_{nn}$\\
 \hline
(1/16) &  $[(\openone-n_\uparrow) (\openone-n_\downarrow)]_i\; [n_\uparrow n_\downarrow]_j$ &0&0 &  \\
 \hline
(1/1) &  $[(\openone-n_\uparrow) (\openone-n_\downarrow)]_i\; [(\openone-n_\uparrow)(\openone-n_\downarrow)]_j$ &0&0 &  \\
 \hline
(6/11) &  $[(\openone- n_\uparrow) n_\downarrow]_i\; [n_\uparrow(\openone-n_\downarrow)]_j$ &0&0 & $G_{nn}$, $G^{z}_{\sigma\sigma}$\\
 \hline
(6/6) &  $[(\openone- n_\uparrow)n_\downarrow]_i\; [(\openone- n_\uparrow)n_\downarrow]_j$ &0&0 & $G_{nn}$, $G^{z}_{\sigma\sigma}$\\
 \hline
(4/13) &  $[-c_\uparrow c_\downarrow]_i\; [c^\dagger_\uparrow c^\dagger_\downarrow]_j$ &-2&0& $G^{(0)'}_p$\\
 \hline
(7/10) &  $[c_\uparrow c^\dagger_\downarrow]_i\; [- c^\dagger_\uparrow c_\downarrow]_j$ &0&-1 & $G^{(+-)}$\\
 \hline
\multicolumn{5}{|c|}{Group-3(algebraic)  $\nu_{g3}=3.89(6)$}\\
 \hline
 \hline
(11/16) &  $[n_\uparrow (\openone- n_\downarrow)]_i\; [n_\uparrow n_\downarrow]_j$ &0&0& $G_{nn}$\\
 \hline
(1/11) &  $[(\openone- n_\uparrow)(\openone- n_\downarrow)]_i\; [n_\uparrow (\openone- n_\downarrow)]_j$ &0&0 &  \\
 \hline
(1/6) &  $[(\openone -n_\uparrow) (\openone- n_\downarrow)]_i\; [(\openone- n_\uparrow) n_\downarrow]_j$ &0&0 &  \\
 \hline
(6/16) &  $[(\openone- n_\uparrow)n_\downarrow]_i\; [ n_\uparrow n_\downarrow]_j$ &0&0 & $G_{nn}$\\
 \hline
\multicolumn{5}{|c|}{Mutual information  $\nu_{I}=1.95(4)$}\\
 \hline
 \hline
\end{tabular}
}
\caption{Similar to Table~\ref{tab:ops-group-s12} but for the
  half-filled SU(2) Hubbard model at $U=0$. 
  Generalized correlation functions 
  $\langle{\cal T}^{(m)}_i {\cal T}^{(n)}_j\rangle$ are denoted as $(m/n)$, 
  and only the $m\leq n$ components are shown.
  Here $\Delta N_c$ and $\Delta S$ indicate how 
  ${\cal T}^{(m)}_i$ changes the charge and spin quantum numbers,
  respectively. 
  The corresponding conventional correlation functions, $G$, are
  listed in the last column. 
}
\label{tab:ops-group-su2u0}
\end{table}
The correlation functions with the slowest algebraic decay are those in group 1.
They are composed of operators describing single-particle transfer
processes weighted by the occupation number of the spin-up and
down-particle in various ways. 
Specifically, these generalized correlation
functions have the form 
$\langle A_iB_i A^\dagger_j B^\prime_j\rangle$,
where, for example, $A\equiv c_1$, and $B$ and $B^\prime\in\{n_2, 1-n_2\}$.
The corresponding single-site transition operators change the number
of electrons by $\Delta N_c=\pm 1$ and the spin by $\Delta S^z=\pm 1/2$,
and we obtain the exponent $\nu_{\rm  g1}\simeq 0.98(4)$.
Operators in the generalized correlation functions
falling into group 2 describe density-like correlations,
$\Delta N_c=0$, $\Delta S^z=0$,
and decay algebraically with an exponent 
$\nu_{\rm g2}\simeq 3.89(6)$. 
The generalized correlation
functions have the form 
$\langle A_iB_i A^\prime_j B^\prime_j\rangle$ 
with $A_i,A^\prime_j,B_i,B^\prime_j\in\{n_1,n_2,(1-n_1),(1-n_2)\}$,
and the corresponding single-site transition operators are related to
basis states with odd numbers of particles.
Operators falling into group 3 describe spin-, density-, and pair-like
correlations 
$\Delta N_c\in\{0,\pm 2\}, \Delta S^z\in\{\pm 1,0\}$,
and decay algebraically with an exponent 
$\nu_{\rm g3}\simeq 1.93(5)$.  
The corresponding single-site transition operators are related to
basis states with even numbers of particles.

We now consider the physical correlation functions, in particular, the
equal-time single- and two-particle Green functions.
According to Eq.~(\ref{eq:G_nn}), $G_{nn}$ can be expressed as a
linear combination of generalized correlation functions from group 2 
and group 3; therefore, the smallest $\nu$ of these two groups determine 
$\nu_{G_{nn}}\simeq 1.92(5)$, which is in agreement with the analytic result  
given in Table~\ref{tab:su2-corr}. 
According to Eqs.~(\ref{eq:G_z}) and ~(\ref{eq:G_xy}), $G^{zz}$ and
$G^{+-}$ are composed of generalized correlation functions
from group 3 only; we obtain $\nu_{G_z}=\nu_{G_{xy}}\simeq 1.92(5)$,
which is also consistent
with the analytic result given in Table~\ref{tab:su2-corr}. 
According to Eq.~(\ref{eq:G_cc}), $G_{c^\dagger_\sigma c_\sigma}$
is composed of generalized correlation functions from group 1 only; we
obtain  
$\nu_{c^\dagger_\sigma c_\sigma}\simeq 0.98(4)$, reproducing the
analytic result given in Table~\ref{tab:su2-corr}.
According to Eq.~(\ref{eq:G_1}), $G_p^{(0)}$ is composed of
generalized correlation functions from group 3 only; we obtain
$\nu_{G_p^{(0)}}=1.92(6)$, in agreement with the analytic
result given in Table~\ref{tab:su2-corr}.  
We have again confirmed that the mutual information 
decays with the exponent $\nu_I=1.95(4)$, which  
decays twice as fast as the most slowly decaying
correlation function, 
which in this case corresponds to one-particle transfer processes.
This means that the quasi-long-range entanglement in the metallic case
is due to one-particle-like hopping, as expected.  

\begin{figure}[!htb]
\centerline{
  \includegraphics[scale=0.43]{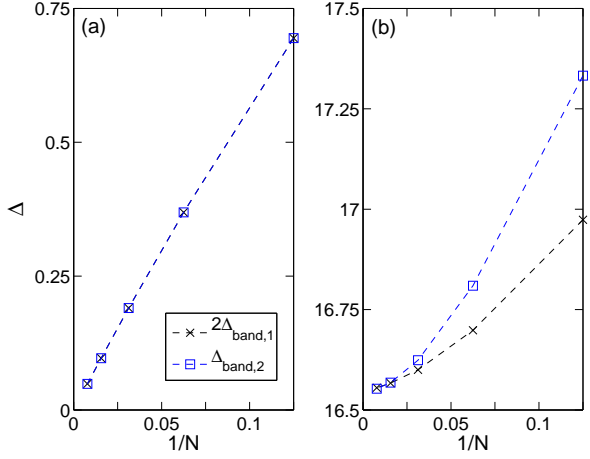}
}
\caption{(Color online) Finite-size scaling of the single-particle
  ($\Delta_{\rm band}$) and pair ($\Delta_{\rm pair}$) gap for the
  half-filled SU(2) Hubbard model at (a) $U=0$ and (b) $U=10$.
}  
\label{fig:su2_gaps_u0}
\end{figure}

For $U=10$, it is known from the analytic picture that the
charge modes are gapped, while the spin modes are
gapless.
Thus, the correlation functions that couple to the charge mode
are expected to decay
exponentially, while those that couple to the spin mode 
are expected to decay
algebraically.
% HERE U=10
\begin{figure}[!bh]
\centerline{
  \includegraphics[scale=0.47]{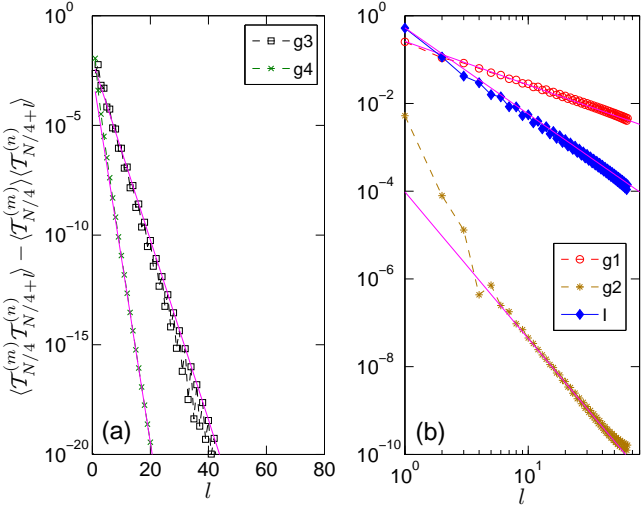}
}
\caption{(Color online)
  Correlation functions 
  measured relative to the middle of the chain 
  for $U=10$, with (a) the two groups of exponentially decaying functions
  plotted on a semi-logarithmic scale and (b) the two groups of
  algebraically decaying correlation functions plotted on a log-log
  scale, both as a function of distance $l$.
  The straight lines are the results of our fits.
}   
\label{fig:su2-correlfuns-u10}
\end{figure}
Classifying the correlation functions 
according to their type of decay and the value of the correlation
length or the exponents
as described above, we 
find two groups of
exponentially decaying and two groups of algebraically decaying
correlation functions,
as summarized in Table~\ref{tab:ops-group-su2u10}.
%HERE appendix
\begin{table}[!htb]
\centering
\centering\scalebox{0.86}{
\begin{tabular}{|c|}
 \hline
 \hline
\multicolumn{1}{|c|}{Group-1(algebraic), $\nu_{g1}=0.98(3)$}\\
 \hline
 (7/10),
 (10/7), 
 (6/11), 
 (11/6),
 (6/6), 
 (11/11)\\ 
 \hline
 \hline
\multicolumn{1}{|c|}{Group-2(algebraic), $\nu_{g2}=3.78(3)$}\\
 \hline
 (11/16), 
 (16/11), 
 (1/11), 
 (11/1), 
 (1/16),
 (16/1),\\ 
 (6/16), 
 (16/6), 
 (1/6), 
 (6/1),
 (1/1), 
 (16/16)\\ 
 \hline
 \hline
\multicolumn{1}{|c|}{Group-3(exponential), $\xi_{g3}=1.07(2)$}\\
 \hline
 (12/15), 
 (15/12), 
 (5/12), 
 (12/5), 
 (3/14), 
 (14/3), 
 (8/14), 
 (14/8),\\
 (2/15), 
 (15/2), 
 (2/5),
 (5/2),
 (3/9), 
 (9/3), 
 (8/9), 
 (9/8)\\
 \hline
 \hline
\multicolumn{1}{|c|}{Group-4(exponential), $\xi_{g4}=0.51(3)$}\\
 \hline
 (4/13), (13/4)\\
 \hline
 \hline
\multicolumn{1}{|c|}{Mutual information(algebraic), $\nu_I=2.08(5)$}\\
 \hline
\end{tabular}
}
\caption{Similar to Table~\ref{tab:ops-group-s12} but for the half-filled SU(2) Hubbard model at $U=10$.
The corresponding single-site transition operators are
summarized in Eq.~(\ref{eq:fermion-ops-1}).
The generalized correlation functions 
$\langle{\cal T}^{(m)}_i {\cal T}^{(n)}_j\rangle$ are denoted as $(m/n)$. 
}
\label{tab:ops-group-su2u10}
\end{table}
We find that the mutual information again decays twice as fast
as the slowest algebraically decaying correlation functions; 
we obtain
an exponent $\nu_I=1.95(5)\simeq 2 \times 0.98(3)$. 
The most slowly decaying correlation functions 
($\nu_{{\rm g} 1}=0.98(3)$) are in
group 1 and correspond to spin-flip excitations, 
$\Delta N_c=0, \Delta S^z\in\{\pm 1,0\}$,
as is to be expected from the spin-density-like
nature of the ground state.
The one- and two-particle hoppings (groups 3 and 4) decay exponentially,
and we find that the decay length of the two-particle hopping is twice of
that of the single-particle hopping.
  The reason for this behavior is that
 the two-particle gap is twice as large
as the one-particle gap in the thermodynamic limit, as is shown
in Fig.~\ref{fig:su2_gaps_u0}(b).
Note that the relation  $\Delta_{\rm pair }(N)=2\times\Delta_{\rm band}(N)$ 
holds for any $N$ at $U=0$, while this becomes true only in the
thermodynamic limit for finite $U$ (Figs.~\ref{fig:su2_gaps_u0}(a),(b)).

By carrying out a similar analysis for the quarter-filled case at $U=10$, we  
obtain the results displayed in Table~\ref{tab:ops-group-su2u10-f14}.
%HERE appendix
\begin{table}[!htb]
\centering
\centering\scalebox{0.86}{
\begin{tabular}{|c|}
 \hline
 \hline
\multicolumn{1}{|c|}{Group-1(algebraic), $\nu_{g1}=1.12(3)$}\\
 \hline
(12/15),  
(15/12),  
(5/12), 
(12/5),  
(3/14), 
(14/3),  
(8/14), 
(14/8),\\ 
(2/15), 
(15/2),  
(8/9), 
(9/8), 
(2/5), 
(5/2), 
(3/9), 
(9/3)\\ 
\hline
\hline
\multicolumn{1}{|c|}{Group-2(algebraic), $\nu_{g2}=1.48(4)$}\\
 \hline
(7/10),  
(10/7),  
(1/11),  
(11/1),  
(6/11),  
(11/6), 
(1/6),  
(6/1),\\  
(1/1), 
(6/6), 
(11/11)\\  
 \hline
 \hline
\multicolumn{1}{|c|}{Group-3(algebraic), $\nu_{g3}=2.47(4)$}\\
 \hline
 (11/16), 
 (16/11), 
 (1/16), 
 (16/1), 
 (6/16),
 (16/6), 
 (16/16),\\ 
 (4/13),
 (13/4)\\ 
\hline
\hline
{Mutual information(algebraic), $\nu_I=2.19$}\\
 \hline
\end{tabular}
}
\caption{Similar to Table~\ref{tab:ops-group-s12} but for the
  quarter-filled SU(2) Hubbard model at $U=10$. 
  The corresponding single-site transition operators are
  summarized in Eq.~(\ref{eq:fermion-ops-1}).
  The generalized correlation functions  
  $\langle{\cal T}^{(m)}_i {\cal T}^{(n)}_j\rangle$ are denoted as $(m/n)$.  
}
\label{tab:ops-group-su2u10-f14}
\end{table}
The generalized correlation functions fall into three groups
with exponents 
$\nu_{\rm g1}= 1.12(3)$,
$\nu_{\rm g2}= 1.48(4)$,
and $\nu_{\rm g3}= 2.47(4)$.
Applying Eqs.~(\ref{eq:G_nn})--(\ref{eq:G_1}), we
obtain the corresponding exponents of  
the conventional correlation functions, 
$\nu_{G_{c^\dagger_\sigma c_\sigma}}= 1.12(3)$,   
$\nu_{G_{nn}} = \nu_{G^{zz}}=\nu_{G^{+-}} = 1.48(4)$, and    
$\nu_{G^{(0)}_{p}} \nu\simeq 2.47(4)$,
again in agreement with 
the analytic results given in
Table~\ref{tab:su2-corr}.
The mutual information decays algebraically with an
exponent $\nu_I=2.19(4)$,
indicating that the 
long-range entanglement is due to the one-particle hopping term 
here as well.
%%%%

\subsection{SU($n$) Hubbard model with $n>2$}

\begin{figure}[!htb]
\centerline{
  \includegraphics[scale=0.6]{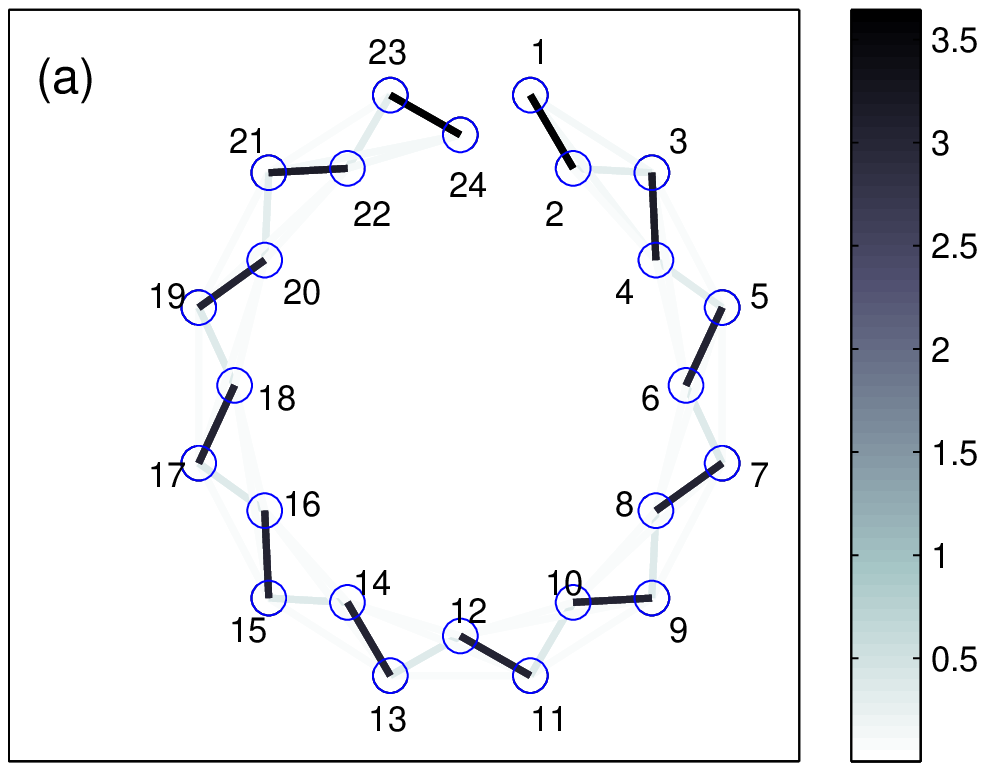}
}
\centerline{
  \includegraphics[scale=0.5]{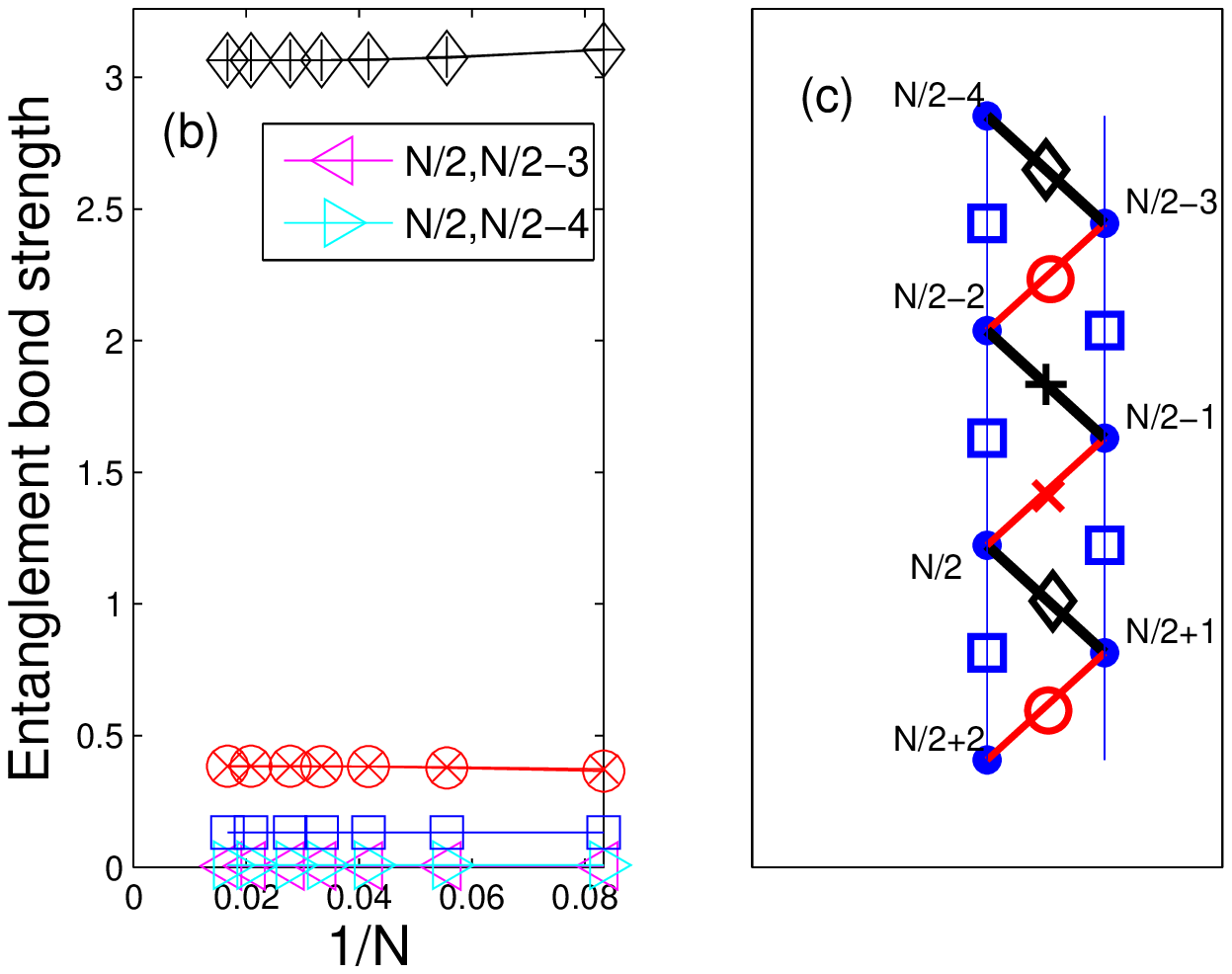}
}
\caption{(Color online) 
    (a) Graphical representation of the components of the two-site
    mutual information for the 
    SU(4) Hubbard model at half filling for $N=24$ lattice sites. 
    (b) Finite-size scaling of the leading entanglement bonds. 
    (c) Entanglement pattern in the thermodynamic limit,
    with the width of the lines indicating
    the strength of the entanglement bonds.
    Entanglement bonds 
    $I_{N/2,N/2-1}$,
    $I_{N/2,N/2+1}$,
    $I_{N/2,N/2-2}$,
    $I_{N/2-1,N/2-2}$,
    $I_{N/2+1,N/2+2}$,
    are denoted by the symbols 
    {\color{red}{$\pmb{\times}$}},  
    {\color{black}{$\pmb{\lozenge}$}},  
    {\color{blue}{$\pmb{\Box}$}}, 
    {\color{red}{$\pmb{\ocircle}$}},  and
    {\color{black}{$\pmb{+}$}}, respectively. 
}    
\label{fig:su4_f12}
\end{figure}

Next we consider the SU($n$) Hubbard model with  
$n>2$ at 
half filling, $f=1/2$, for large Coulomb interaction, $U=10$. 
It has been shown using bosonization that, in this case, the
ground state is dimerized for even values of $n$.\cite{marston}
The same scenario has also been shown numerically to occur for odd $n$
values.\cite{szirmai-2008}
For $n=3$ and $q<n$, only the half-filled case occurs; therefore,  
we investigate the entanglement pattern for
the $n=4$ and $n=5$ cases only. The results for a  
a finite chain of length $N=24$ for $n=4$ are displayed
in Fig.~\ref{fig:su4_f12}(a), finite-size scaling of the
leading entangled bonds in Fig.~\ref{fig:su4_f12}(b), and the
resulting entanglement bonds in the thermodynamic limit in
Fig.~\ref{fig:su4_f12}(c). 
It can be clearly seen that the strongly and weakly entangled bonds alternate 
along the chain, as is expected for a dimerized phase.

\begin{figure}[!htb]
\centerline{
  \includegraphics[scale=0.5]{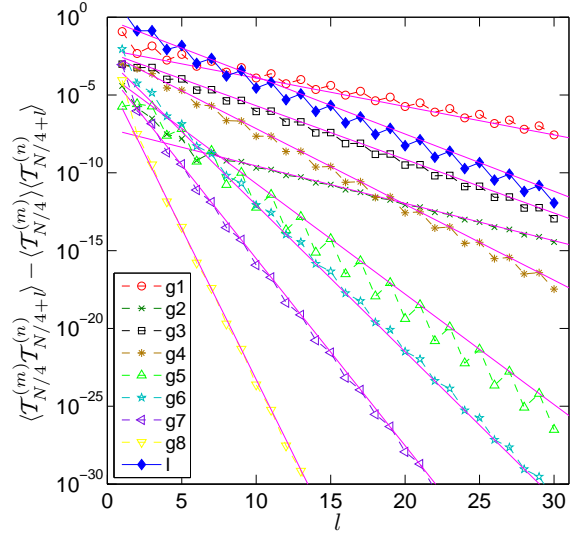}
}
\caption{(Color online)
Similar to Fig.~\ref{fig:su2-correlfuns-u10} but 
for the half-filled SU(4) Hubbard model with 
$N=60$ sites for $U=10$.}
\label{fig:su4f12-correlfuns-u10}
\end{figure}

Since the spin and charge modes are coupled and 
both modes are gapped, all generalized correlation
functions decay exponentially, 
$G_{i,i+l} =\langle A_iB_{i+l}\rangle \sim \exp(-l/\xi)$ with $\xi$
the decay length, 
as can be seen in Fig.~\ref{fig:su4f12-correlfuns-u10}.
We group the generalized correlation functions into groups with the
same decay length and find that they fall into eight groups, 
tabulated in Table~\ref{tab:ops-su4-f12}.
\begin{table}
\centering\scalebox{0.86}{
\begin{tabular}{|c|c|c|c|c|}
 \hline \hline
 $\xi$ & group  & $\Delta N_c$ & $\Delta S^z$ & Representative operator \\ 
       & number &              &              &  combination: $\langle{\cal T}^{(m)}_i {\cal T}^{(n)}_j\rangle$\\
\hline
\hline
2.32(4) & g1 & 0 &  $\{\pm 3, \pm 2, \pm 1\}$ & $(c^\dagger_1c_2n_3 n_4)_i\; (c_1 c^\dagger_2 n_3n_4)_j$ \\%& 1 hop left and 1 hop right OK  1-1\\
    & g1 & 0 &  0 & $(n_1n_2n_3(1-n_4))_i\; \times$\\
    &    &   &    & $(n_1n_2(1-n_3)n_4)_j$ \\%some kind of density
1.92(3) & g2 & 0 & $\{\pm 4, \pm 2, 0\}$ & $(c^\dagger_1c^\dagger_2c_3c_4)_i\; (c_1c_2c^\dagger_3c^\dagger_4)_j$ \\%& 2-right and 2-left hopping, or only density OK 2-2\\
        & g2 & 0 & 0 & $(n_1n_2n_3n_4)_i\; (n_1n_2n_3n_4)_j$ \\%some kind of density
1.25(4) & g3 & $\pm 1$ & $-\frac{3}{2},\dots,\frac{3}{2}$ & $(c^\dagger_1n_2n_3 n_4)_i\; (c_1 n_2 n_3 n_4)_j$ \\%& 1-particle hopping OK 1-0\\
0.88(6) & g4 & $\pm 1$ & $-\frac{7}{2},\dots,\frac{7}{2}$ & $(c^\dagger_1c^\dagger_2c_3n_4)_i\; (c_1c_2c^\dagger_3 n_4)_j$ \\%& 2-part left hopping, 1 part right hopping 2-1\\
0.61(8) & g5 & $\pm 2$ & $-2,\dots,2$ & $(c^\dagger_1 c^\dagger_2 n_3 n_4)_i\; (c_1 c_2 n_3 n_4)_j$ \\%& 2-right hopping OK 2-0\\
0.46(5) & g6 & $\pm 2$ & $\{\pm 3,\pm 1\}$ & $(c^\dagger_1c^\dagger_2c^\dagger_3c_4)_i\; (c_1c_2c_3c^\dagger_4)_j$ \\%& 3-particle left, 1 particle right hopping OK 3-1\\
0.36(7) & g7 & $\pm 3$ & $-\frac{3}{2},\ldots,\frac{3}{2}$ & $(c^\dagger_1c^\dagger_2c^\dagger_3 n_4)_i\; (c_1c_2c_3n_4)_j$ \\%& 3-particle hop left and 0 right OK 3-0\\
0.22(7) & g8 & $\pm 4$ & 0 & $(c^\dagger_1c^\dagger_2c^\dagger_3c^\dagger_4)_i\; (c_1c_2c_3c_4)_j$ \\%& 4-particle hopping OK 4-0\\
\hline
1.16(4)  & I & & & Mutual Information \\
\hline
\hline
\end{tabular}
}
\caption{Categorization of the generalized correlation functions for the half-filled SU(4) Hubbard model at $U=10$.}
\label{tab:ops-su4-f12}
\end{table}
Since the generalized correlation functions describe all allowed
transitions from an initial to a final state on a subsystem 
consisting of two sites,  
they encompass processes which correspond to spin and
density correlations, as well as one- and multi-particle hopping on
those two sites.
The operators $A_i$ in the correlation functions can now be characterized 
by eight independent parameters or by their proper symmetric 
and antisymmetric combinations.
In analogy to results of conformal field theory,
we expect that the decay lengths might be some simple  
function of the eight parameters.
The possible values of $\Delta N_c$ and $\Delta S^z$ 
corresponding to the eight different groups are 
given in Table~\ref{tab:ops-su4-f12}. 
In general, the single-site transition operators, 
${\cal T}_i^{(m)}$, can 
contain $r$ fermion operators 
and $4-r$ density-like operators with $r=0,\dots,4$. In addition, the
$r$ fermion operators 
are constructed from various combinations of creation and annihilation
operators 
with different spin values.
Some representative operator combinations are listed in the last column of
Table~\ref{tab:ops-su4-f12}.
It is clear that the largest decay length corresponds to those generalized 
correlation functions where $A_i$ does not change the particle number,
i.e., $\Delta N_c=0$. 
These are the slowest decaying correlation functions and are collected
in groups 1 and 2.  
Group 2 also includes single-site transition operators with $r=0$
fermion operators,   
and they are related to components of spin- and density-like
conventional correlation functions.
In the subsequent groups, as the single-site transition operator adds
more and more particles  
to the system, the corresponding decay length decreases. 
The fastest decaying correlation functions correspond to
$\Delta N_c=4$, which 
is the maximum value. 
In addition to the strong $\Delta N_c$ dependence,
the decay length also depends on $\Delta S^z$ 
because the related single-site transition operators describe different
spin-flip processes. 
Single-site transition operators in group 3, 5, 7, and 8 contain $r=1$, 2, 3, 
and 4 fermionic creation operators and $4-r$ density-like operators,
respectively. 
Their corresponding decay length drops systematically by a factor of
two within our numerical error. 
A similar tendency is observed for groups 2, 4, and 6.
It is worth reiterating that the decay length  
of $\langle {\cal T}_i^{m}{\cal T}_j^{n}\rangle$ depends only 
on the number of creation and annihilation operators, 
i.e., it is independent of the various combinations of the 
density-like operators by which the action of the fermion operators
are weighted.

\begin{figure}[!htb]
\centerline{
  \includegraphics[scale=0.43]{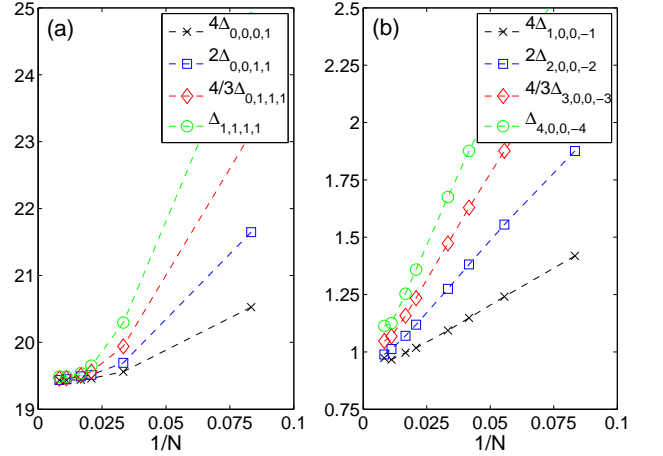}
}
\caption{(Color online) Finite-size scaling of (a) the band gaps related
  to adding one, two, three, or four 
  particles to the system and (b) the spin gap 
  for the half-filled  
  SU(4) Hubbard model at $U=10$.   
  The gaps are denoted by 
  $\Delta_{\Delta N_1,\Delta N_2,\Delta N_3,\Delta N_4}$, where the labels 
  ${\Delta N_\sigma}$ show the change in the number of particles with
  spin $\sigma$.
} 
\label{fig:su4_gaps_f12_u10}
\end{figure}

In order to further elucidate the origin of the behavior of the decay
lengths of the
different groups, we have calculated the band gaps formed by adding
particles with different flavors 
as well as by adding particles with
the same flavor.
In both cases, the scaled gaps become equal in the thermodynamic limit,
indicating that the level structure in the energy spectrum is
uniformly spaced. 
A similar structure has been found in the spin sector as well.
This behavior is shown for the half-filled SU(4)
Hubbard model in Fig.~\ref{fig:su4_gaps_f12_u10}. 
The fact that higher excitation gaps are integer multiples of
the one-particle gap holds already for the finite chain with $N=60$ lattice
sites for which the decay lengths of the generalized correlation functions
have been determined. These decay lengths are sometimes but not always
integer multiples of each other. 
In addition, the observed trend
for groups 3, 5, 7, and 8 and for groups 2, 4, and 6 is not linear.
This can be understood by the fact that when band gaps are determined,
the particles are added to the overall system.
Thus, they are delocalized along the whole chain,
whereas, for single-site transition
operators with increasing $\Delta N_c$,
more and more particles are added to a single-site subsystem. 
Higher excitations are possible in the latter case, so the decay
length can decrease faster 
than the band gaps for the same $\Delta N_c$ values.

In summary, the slowest decaying generalized correlation functions
belong to group 1, which have a decay length of $\xi_{g1}=2.32$.
The two-site mutual information also decays exponentially
with a decay length $\xi_I=1.16=\xi_{g1}/2$. 
Therefore, it has half of the longest decay length of the slowest
decaying generalized correlation function. 
%%%

\begin{figure}[!t]
\centerline{
  \includegraphics[scale=0.6]{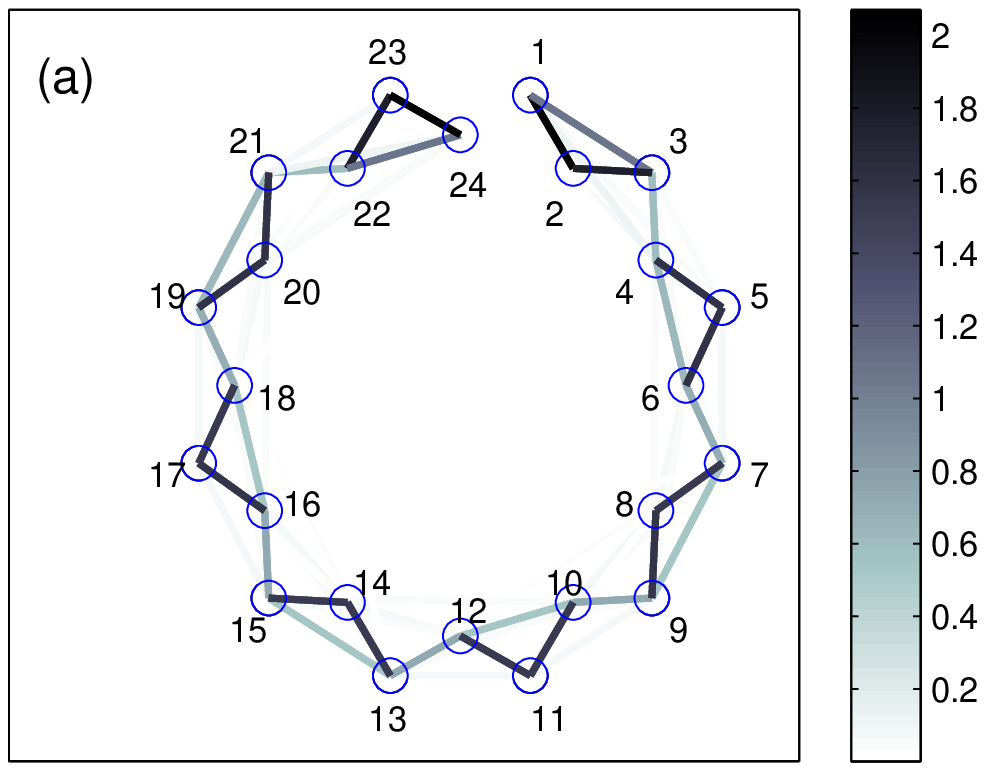}
}
\centerline{
  \includegraphics[scale=0.5]{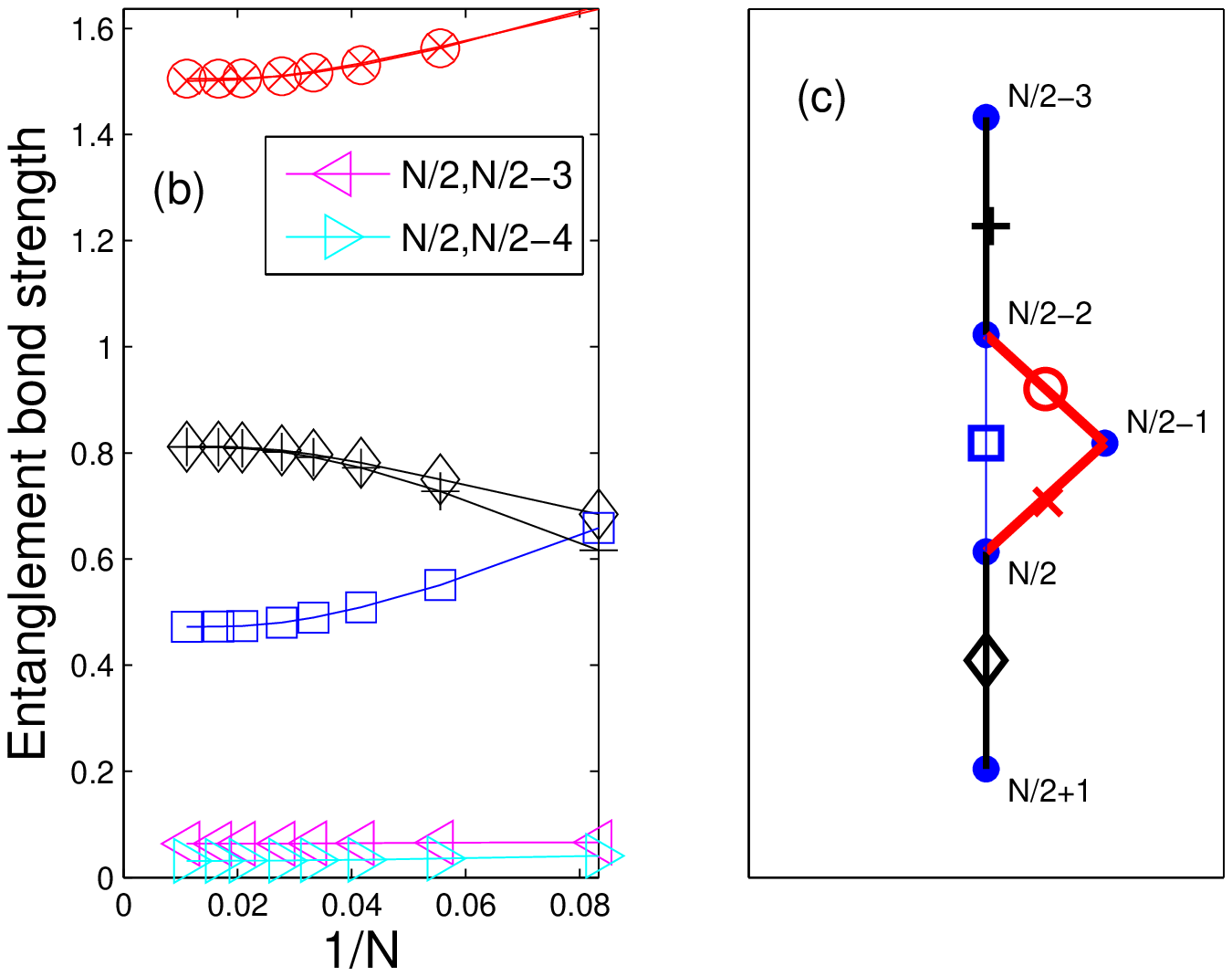}
}
\caption{(Color online) Similar to Fig.~\ref{fig:su4_f12} but for the
  SU(4) Hubbard model at one-third-filling. 
    The entanglement bonds 
    $I_{N/2,N/2-1}$,
    $I_{N/2,N/2+1}$,
    $I_{N/2,N/2-2}$,
    $I_{N/2-1,N/2-2}$,
    $I_{N/2-2,N/2-3}$
    are denoted by the symbols 
  {\color{red}{$\pmb{\times}$}},  
  {\color{black}{$\pmb{\lozenge}$}},  
  {\color{blue}{$\pmb{\Box}$}}, 
 {\color{red}{$\pmb{\ocircle}$}}, and
 {\color{black}{$\pmb{+}$}}, respectively.
} 
\label{fig:su4_f13}
\end{figure}

\begin{figure}[!t]
\centerline{
  \includegraphics[scale=0.45]{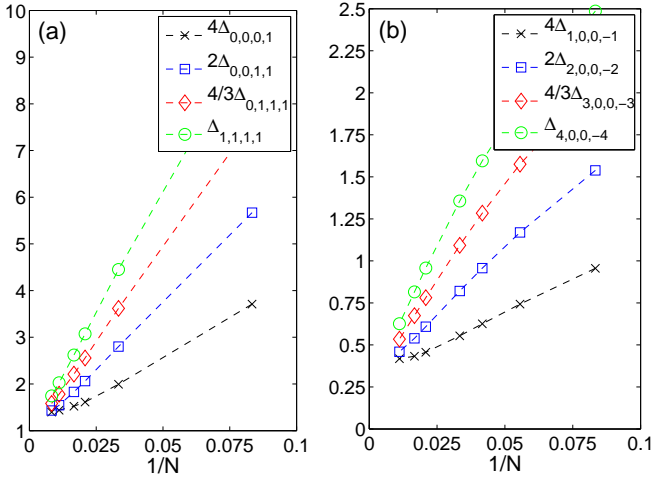}
}
\caption{(Color online) Similar to Fig.~\ref{fig:su4_gaps_f12_u10} but for
the one-third-filled SU(4) Hubbard model at $U=10$.}
\label{fig:su4_gaps_f13_u10}
\end{figure}

\begin{figure}[!t]
\centerline{
  \includegraphics[scale=0.5]{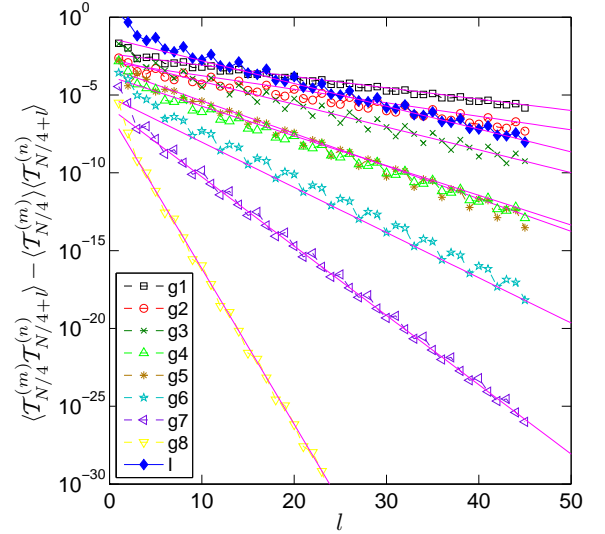}
}
\caption{(Color online) Similar to Fig.~\ref{fig:su2-correlfuns-u10},
  but for the SU(4) Hubbard model at $f=1/3$ filling and $U=10$ 
  calculated for a chain with $N=90$ sites.}
\label{fig:su4f13-correlfuns-u10}
\end{figure}

A similar analysis for $f=1/3$, i.e., the one-third-filled case, is
presented in Fig. ~\ref{fig:su4_f13}. 
Note that highly entangled three-site units are
connected by a single entanglement bond.
The strengths of the extrapolated values of the entanglement bonds
show that next-nearest-neighbor entanglement  
is also important within the three-site entangled units.
In addition, the entanglement map has left-right symmetry with respect 
to the mid-point of the three-site unit.
The characterization of the correlation functions is summarized in
Table~\ref{tab:ops-su4-f13}, 
and finite-size scaling of the relevant energy gaps is shown in 
Fig.~\ref{fig:su4_gaps_f13_u10}. 
The scaled gaps become equal again in the thermodynamic limit,
indicating an equidistant level structure in the energy spectrum of
both the charge and spin sectors.
According to Fig.~\ref{fig:su4f13-correlfuns-u10}
the decay length of the slowest decaying correlation function is
$\xi_{g1}=6.18(5)$,
for which $\Delta N_c=0$ and $\Delta S^s=0$. 
The decay length of the two-site mutual information 
is $\xi_I=3.02(4)=\xi_{g1}/2$.
\begin{table}
\centering\scalebox{0.87}{
\begin{tabular}{|c|c|c|c|c|}
 \hline \hline
 $\xi$ & group  & $\Delta N_c$ & $\Delta S^z$ & Representative operator \\
       & number &              &              & combination: $\langle{\cal T}^{(m)}_i {\cal T}^{(n)}_j\rangle$\\
% \hline\multicolumn{4}{|c|}{Group-1(algebraic)}\\
\hline
\hline
6.18(5) & g1 & $\pm 1$ & $-\frac{3}{2},\dots,\frac{3}{2}$  & $(c^\dagger_1n_2n_3n_4)_i\; (c_1n_2n_3n_4)_j$ \\%& 1-particle hopping \\
4.54(4)  & g2 & 0 &  $\{\pm 3, \pm 2, \pm 1\}$ & $(c^\dagger_1c_2n_3n_4)_i\; (c_1c^\dagger_2n_3n_4)_j$ \\%& blabla \\ 1-particle left 1-right hopping
        & g2 & 0 & 0 & $(n_1n_2n_3n_4)_i\; (n_1n_2n_3n_4)_j$ \\
3.03(3) & g3 & $\pm 1$ & $-\frac{7}{2},\ldots,\frac{7}{2}$ & $(c^\dagger_1c^\dagger_2c_3n_4)_i\; (c_1c_2c^\dagger_3n_4)_j$ \\%& 2-right,2left hopping \\

2.27(3) & g4 & $\pm 2$ & $-2,\dots,2$  & $(c^\dagger_1c^\dagger_2n_3n_4)_i\; (c_1c_2n_3n_4)_j$ \\%& 2-part right hopping \\
%        & g4 & 0 & $\{\pm 4, \pm 2, 0\}$ & $(c^\dagger_1c^\dagger_2c_3c_4)_i\; (c_1c_2c^\dagger_3c^\dagger_4)_j$ \\%& 2-right and 2-left hopping, or only density OK 2-2\\

2.04(7) & g5 & 0 & $\{\pm 4,\pm 2, 0\}$ & $(c^\dagger_1c^\dagger_2c_3c_4)_i\; (c_1c_2c^\dagger_3c^\dagger_4)_j$ \\

1.49(3) & g6 & $\pm 2$ & $\{\pm 3,\pm 1\}$  & $(c^\dagger_1c^\dagger_2c^\dagger_3c_4)_i\; (c_1c_2c_3c^\dagger_4)_j$ \\%& 3-particle left 1 right hopping \\

1.0(2) & g7 & $\pm 3$ & $-\frac{3}{2},\ldots,\frac{3}{2}$  & $(c^\dagger_1c^\dagger_2c^\dagger_3n_4)_i\; (c_1c_2c_3n_4)_j$ \\%& 3-right hopping \\

0.45(4) & g8 & $\pm 4$ & 0  & $(c^\dagger_1c^\dagger_2c^\dagger_3c^\dagger_4)_i\; (c_1c_2c_3c_4)_j$ \\%& 4-particle right hopping \\

\hline
3.02(4) & I & & &  Mutual Information\\%& mutual info \\
\hline
\hline
\end{tabular}
}
\caption{Similar to Table~\ref{tab:ops-su4-f12} but for the
  one-third-filled SU(4) Hubbard model at $U=10$.}
\label{tab:ops-su4-f13}
\end{table}

\begin{figure}[!t]
\centerline{
  \includegraphics[scale=0.6]{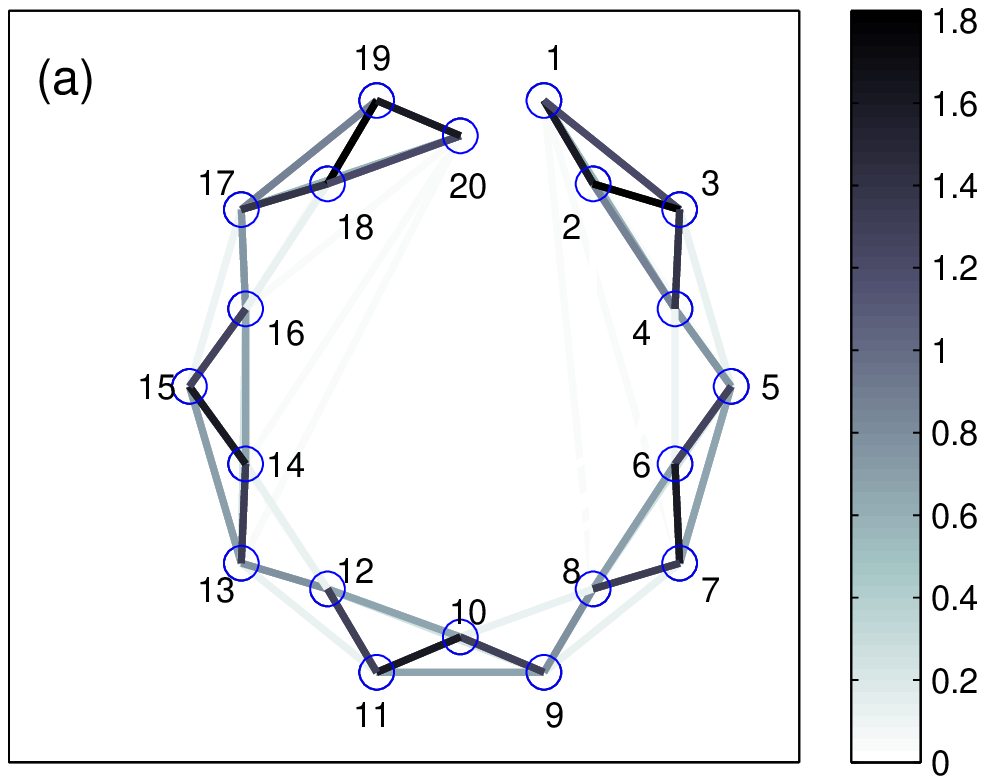}
}
\centerline{
  \includegraphics[scale=0.5]{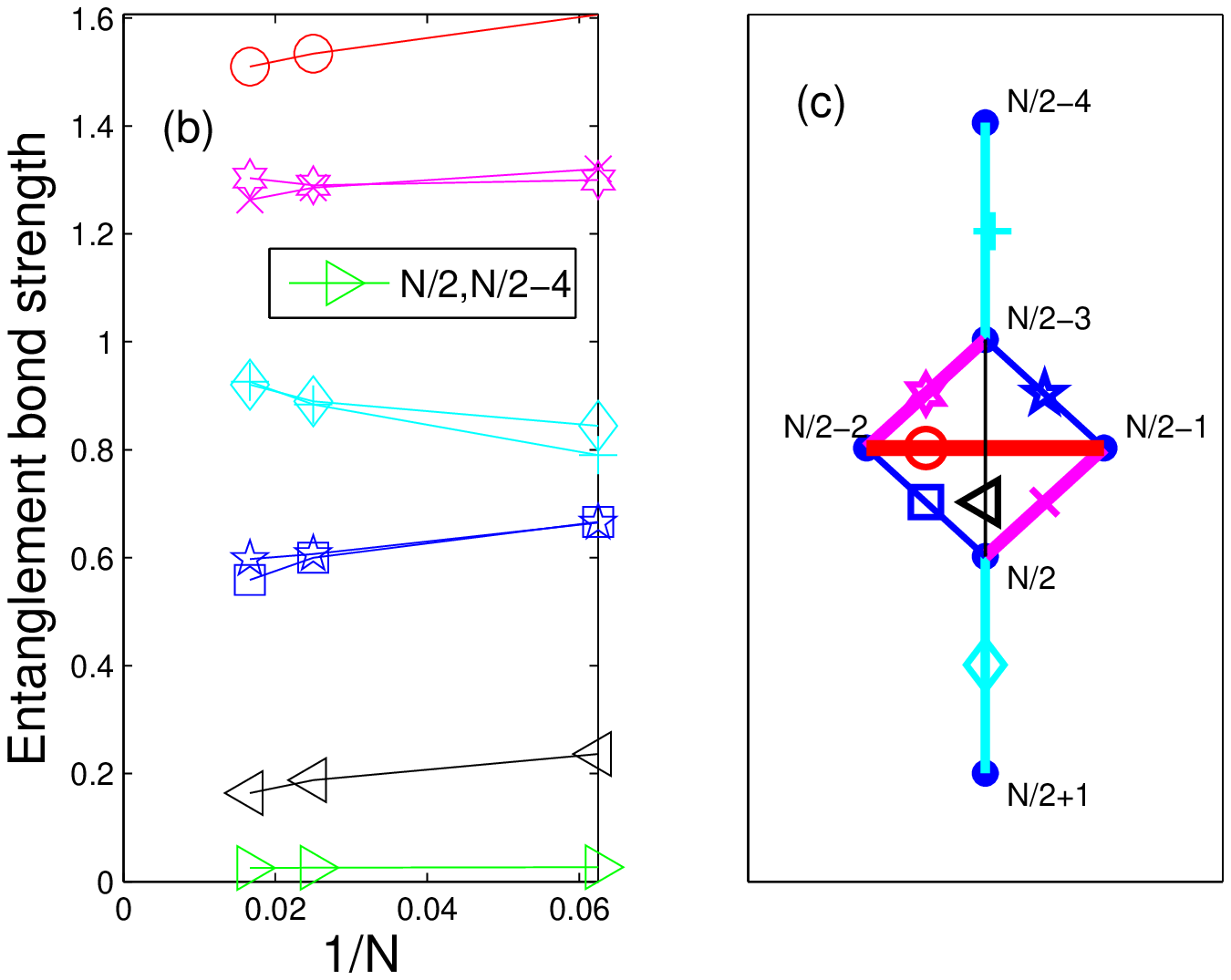}
}
\caption{(Color online) Similar to Fig.~\ref{fig:su4_f12} but for the
  SU(5) Hubbard model at quarter-filling.
    Entanglement bonds 
    $I_{N/2,N/2-1}$,
    $I_{N/2,N/2+1}$,
    $I_{N/2,N/2-2}$,
    $I_{N/2-1,N/2-2}$,
    $I_{N/2-1,N/2-3}$,
    $I_{N/2-2,N/2-3}$
    $I_{N/2-3,N/2-4}$
    are shown by symbols 
 {\color{magenta}{$\pmb{\times}$}}, 
  {\color{cyan}{$\pmb{\lozenge}$}},  
  {\color{blue}{$\pmb{\Box}$}}, 
 {\color{red}{$\pmb{\ocircle}$}},  
 {\color{blue}{$\pmb{\textbf{\ding{73}}}$}}, 
 {\color{magenta}{$\pmb{\davidsstar}$}},   
  {\color{cyan}{$\pmb{+}$}}, respectively. 
} 
\label{fig:su5_f14}
\end{figure}

We consider the case of higher flavor number, $n=5$, in
order to study even more extended spatial entanglement patterns.
For $f=1/4$, i.e., the quarter-filled case, a quadrimerized phase
occurs, as can be seen in Fig.~\ref{fig:su5_f14}.
The highly entangled four-site units are connected by single 
entanglement bonds. The strengths of the 
extrapolated values of the entanglement bonds indicate
that next-nearest-neighbor entanglement is present within the
four-site structure. In addition,  second- and
third-nearest-neighbor entanglement bonds also remain finite
in the thermodynamic limit.

\section{Conclusion}
\label{sec:conclusion}

In this paper, we have formulated a method to characterize
entanglement patterns in correlated systems.
The measure that we use is the two-site mutual information, which is
formed from the von Neumann entropy of the two-site and one-site
density matrices.
We have shown that 
the reduced density matrix of two- and three-site
subsystems can be expressed in terms of two- and three-site generalized
correlation functions related to all two- and three-site
correlation functions based on site-local operators.
This procedure can be extended to $n$-site subsystem; thus  
the mutual information can be generalized to the $n$-site case.
The major part of this work, in particular, the numerical
calculations, concentrate on the two-site case.

For the $S=1/2$ Heisenberg model, we have shown explicitly how the
generalized correlation functions are constructed and have
demonstrated that the dominant long-distance behavior of the mutual
information follows the square of the
most slowly decaying correlation function, which decays algebraically
as the inverse of the distance.
For the anisotropic Heisenberg model, 
both the
generalized and the physical correlation functions can be divided into
two groups that decay algebraically with different 
exponents. 
These exponents can be understood in terms of the
known exact values 
for the transverse and 
longitudinal components of the spin correlations.

In Sec.\ \ref{sec:su_n}, we have applied the methods
to the one-dimensional SU($n$)-symmetric
generalized Hubbard model.
For the half- and quarter-filled $n=2$ case, we classify the
generalized correlation functions into groups according to type of
decay (algebraic or exponential) and rapidity of decay (value of
exponent and correlation length, respectively).
The classification is consistent with the known excitation spectra at
these points.
We find that the two-site mutual information again decays as the
square of the most slowly decaying correlation functions at long
distances.

For higher spin models, i.e., $n=3$ at half filling for large Coulomb
interaction, we have confirmed the expected entanglement pattern of 
the dimerized ground state in which strong and weak bonds
between nearest-neighbor pairs alternate on successive bonds.
For $n=4$ at one-third-filling, highly entangled three-site units are
connected by a single entanglement bond.
Within such three-site units, next-nearest-neighbor entanglement is also 
important. 
For an even higher number of flavors, $n=5$, 
and at one-third-filling, we have again found 
three-site units with similar properties.  
On the other hand, at quarter filling,
we have found highly entangled four-site units which are connected by 
single entanglement bonds.
Within such entangled four-site
units, second and third nearest-neighbor entanglement bonds
also remain finite in the thermodynamic limit.

Therefore, based on our numerical results, we conclude that 
for $f=1/q$-filling and $q<n$, highly entangled
$q$-site units are connected by single entanglement bonds 
in the ground state and  
within such entangled $q$-site units, first-, second-, and up to
$q-1$-neighbor entanglement bonds also acquire finite strength in the
thermodynamic limit.
In addition, due to gaps in all excitation branches, all 
generalized correlation functions, together with the two-site mutual
information, decay exponentially.
Based on the decay length, they fall into several groups.
We have found that their decay length depends on how the
corresponding single-site  
transition operator changes the number of particles and the spin,
consistent with the picture obtained from
conformal field theory.
The two-site mutual information 
again decays as the square of the most slowly decaying correlation
functions; these correlation functions can also be used to identify
the elementary excitation gap.

We remark that our method can easily be incorporated into a wide class
of matrix-product-state- and tensor-network-state-based methods, of
which the DMRG method is just one example.
In particular, the method can be applied to higher-dimensional lattices
to provide the basis for an optimization procedure
to order matrix-product states and to construct efficient network
topologies for  tensor-network-state approaches.\cite{murg2014}
Furthermore, our procedure can be utilized in other numerical methods
in which one- and two-site operator expectation 
values are directly available.
This is the case, for example, in the Configuration Interaction (CI) 
and Quantum Monte Carlo (QMC) techniques.
Extension of the present work to study three-body entanglement is part
of our planned future work.

\begin{acknowledgements}
We thank Sz.\ Szalay, G.\ Ehlers, and F.\ Gebhard for useful discussions.
This work was supported in part by
Hungarian Research Fund (OTKA) through Grant Nos.~K100908 and ~NN110360. 
{\"O}.L.\ acknowledges support from the Alexander von Humboldt
foundation and from ETH Zurich
during his time as a visiting professor,
and R.M.N.\ acknowledges support from the Deutsche
Forschungsgemeinschaft (DFG) through grant
no.\  NO 314/5-1 in Research Unit FOR 1807.

\end{acknowledgements}
%}

\end{document}